\documentclass[11pt]{article}
 \normalbaselines
\usepackage{enumerate}

\usepackage{dsfont}
\usepackage{fullpage}
\usepackage{amsmath,amsthm,verbatim,amssymb,amsfonts,amscd, graphicx,color}
\usepackage{mathtools}
\newtheorem{theo}{Theorem}[section]

\newtheorem{rem}[theo]{Remark}

\newcommand{\bd}{\begin{displaymath}}
\newcommand{\ed}{\end{displaymath}}
\newcommand{\be}{\begin{equation}}
\newcommand{\ee}{\end{equation}}
\newcommand{\bea}{\begin{eqnarray}}
\newcommand{\eea}{\end{eqnarray}}
\newcommand{\bda}{\begin{eqnarray*}}
\newcommand{\eda}{\end{eqnarray*}}
\newcommand{\ba}{\begin{array}}
\newcommand{\ea}{\end{array}}
\newcommand{\bal}{\begin{align}}
\newcommand{\eal}{\end{align}}
\newcommand{\bae}{\begin{align}}
\newcommand{\eae}{\end{align}}

\newcommand{\B}{I\kern -.35em B}
\newcommand{\R}{I\kern -.35em R}

\newcommand{\mt}{\mapsto}

\newcommand{\D}{{\cal D}}

\newcommand{\e}{\varepsilon}
\newcommand{\al}{\alpha}
\newcommand{\ph}{\varphi}
\newcommand{\sth}{ \, :\;}

\newcommand{\dd}{\mbox{\rm\,d}}

\def\ra{\rangle}

\newcommand{\cR}{{\cal R}}

\def\R{\mathbb{R}}
\newcommand{\reff}{\eqref}
\newcommand{\bino}{\bigskip\noindent}

\renewcommand{\ra}{\rightarrow}

\newcommand{\ddd}{{\bf d}}

\usepackage{graphicx}
\usepackage{natbib}

\begin{document}

\title{\Large \bf A dual-age structured epidemiological model with waning immunity and reinfection}
\author{Raimund M. Kovacevic$^{1}$, Nikolaos  I. Stilianakis$^{2,3}$, Vladimir M. Veliov$^1$}

\date{}
\maketitle

$^1$ Institute of Statistics and Mathematical Methods in Economics, Vienna University of Technology, Wiedner Hauptstraße 8, 1040, Vienna, Austria 

$^2$ Joint Research Centre (JRC), European Commission, Via Enrico Fermi 2749, 21027 Ispra, Italy

$^3$ Department of Biometry and Epidemiology, University of Erlangen-Nuremberg,  Waldstrasse 6, 91054 Erlangen, Germany


$^*$Corresponding author: Raimund.Kovacevic@tuwien.ac.at

\maketitle

\vspace{5pt}
\noindent

\begin{abstract}
Waning immunity and reinfection are critical features of many infectious diseases, but epidemiological models often fail to capture the intricate interaction between an individual's history of immunity and their current infection status; when they do, the approach is usually overly simplistic. We develop a novel dual-age structured model that simultaneously tracks immunity age (time since the last recovery from infection) and infection age (time since infection) to analyze epidemic dynamics under conditions of waning immunity and reinfection. The model is formulated as a system of age-structured partial differential equations that describe susceptible and infected populations stratified by both immunity and infection ages. We derive basic reproduction numbers associated with the model and numerically solve the system using a second-order Runge-Kutta scheme along the characteristic lines.
We further extend the model to explore vaccination interventions, specifically targeting individuals according to their immunity age. Numerical results reveal that higher contact rates produce larger amplitude oscillations with longer inter-epidemic periods. The relationship between initial infection levels and long-term epidemic behavior is nonmonotonic. Vaccination efficiency depends critically on the pathogen load profile across immunity and infection age, with more pronounced pathogen load distributions requiring higher vaccination rates for disease elimination. Most efficient vaccination strategies begin with intermediate immunity ages rather than targeting only fully susceptible individuals. The structured dual-age framework provides a flexible approach to analyzing the dynamics of reinfection and evaluating targeted vaccination strategies based on the history of immunity.
\

\end{abstract}
\vspace{1em}
\section{Introduction} \label{SIntro}

Infectious diseases with waning immunity and the potential for reinfection present significant challenges to public health planning and epidemic control. Waning immunity is a key characteristic of infectious diseases such as SARS-CoV-2 and influenza (\cite{Goldberg+2022} \cite{Fonville+2014,Srivastava+2024,Lindsey+2025}). The duration of immune protection can vary greatly, affecting the long-term effectiveness of vaccination strategies (\cite{Bobrovitz+2023, Wolthuis+2017}). Immunity levels differ between individuals and depend, among other factors, on the time elapsed since the last infection, which is linked to waning immunity (\cite{Krammer-2019}). A gradual loss of immunity extends the period during which a person can be reinfected (\cite{Sanches_de_Prada+2024}).
The burden of disease in the population is influenced by the durability of immunity after natural infection or vaccination against reinfection (\cite{Vattiato+2022}). The protective effect of natural infection against reinfection is marked by complex interactions between host immunity and pathogen evolution, leading to various patterns of reinfection (\cite{Chamaitelly+2025}).

Therefore, a deeper understanding of the dynamics of diseases at the population-level requires mathematical models that explicitly account for both the temporal progression of infection within individuals and the gradual waning of immunity following recovery.

Epidemiological models fail to adequately capture the interplay between an individual's immunity history and their current infection status. While models that consider age-dependent immune protection or transient immunity exist \cite{Ghosh+2025}, \cite{Ghosh+Volpert-23}, \cite{Angelov+2024}, they typically do not simultaneously track both the history of individuals' immunity progression and their current infection stage. 

Several models have explicitly addressed the issue of waning immunity using ordinary differential equations (ODE, e.g. \cite{KHAN2024657} and \cite{NISHIMURA2023113426}) and partial differential equations (PDEs). Models that consider exposure to an infectious agent leading to reinfection, and consequently an immune system boost, have also been developed. Structured models, which explicitly track the time elapsed since infection or recovery, capture the within-host dynamics more accurately than ODEs. These models allow for greater flexibility in defining delay distributions and can potentially reproduce more complex dynamics (\cite{Barbarossa+2018}, \cite{Diekmann+2018,Okuwa+2021, Yang+2024}).
\cite{Diekmann+2018} developed a mathematical model to describe how immune status is distributed, shaped by continuous waning and occasional boosting. Barbarossa and Röst (2015) proposed a model in which recovered individuals are classified by their immunity levels and offered extensions related to the decline of immunity and immunity after reinfection (Barbarossa et al. 2017; 2018). \cite{Yang+Nakata-2021} examined the specific features of a model with waning immunity and boosting, such as the uniqueness of the endemic equilibrium. Similarly, Okuwa et al. (2021) developed an age-structured epidemiological model that accounts for boosting and waning immunity, specifically examining conditions to determine the direction of bifurcation from disease-free to endemic steady states.
In a different approach, Yang et al. (2024) developed an immuno-epidemiological model on complex networks, integrating ordinary differential equations and integral equations. This model connects within-host and between-host dynamics by positing that the mortality caused by the disease is an increasing function of the pathogen load within the host. They demonstrated the existence and stability of equilibrium, completely determined by the basic reproduction number of the between-host system. Simulations indicated that non-exponential distributions and network topology play significant roles in predicting epidemic patterns.
In an integro-PDE model that explicitly considers immunity acquired through infection or vaccination, and the subsequent waning of individuals with heterogeneous immunity levels, Angelov et al. (2024) presented mathematical features such as the existence of a solution and asymptotic behavior. In addition, they explored an optimal vaccination policy problem based on the model.

The present paper introduces a dual-age structured approach that  explicitly tracks individuals through two temporal variables: immunity age $\tau$ (time elapsed since last recovery from infection) and infection age $\theta$ (time elapsed since current infection onset).
The model recognizes that an individual's epidemiological state is fundamentally defined by two histories. Susceptible individuals are indexed by their immunity age $\tau$, reflecting how recently they recovered and thus their current level of immune protection. Infected individuals are indexed by both their infection age $\theta$, which determines their pathogen load and infectiousness, and their immunity age at the time of infection $\tau$, which influences disease progression and recovery dynamics. 

The dual-age structure gives rise to a system of two partial differential equations, featuring nonlocal terms in both the governing equations and the boundary conditions. Individuals age at a constant rate in their respective age coordinates, while the compartments of susceptible and infected individuals are coupled through nonlocal boundary conditions that account for recovery and infection fluxes.

In addition, the model incorporates several key biological features: susceptibility $\sigma(\tau)$ that may increase with immunity age as immunity wanes; pathogen load $v(\theta,\tau)$ and infectiousness that vary with both infection age and prior immunity status; recovery rate $\rho(\theta,\tau)$ and mortality rate $\mu(\theta,\tau)$ that depend on infection progression and immunity history; and contact rates $c(\theta,\tau)$ for infected individuals that may vary with disease stage and immunity background.

This study has three primary objectives. First, we derive the mathematical formulation of the dual-age structured model, obtain solution representations, and implement an efficient numerical simulation method by adapting to the problem a second-order Runge-Kutta discretization scheme. Second, we investigate disease persistence and invasion dynamics by computing the relevant reproduction numbers through two complementary approaches: the next-generation matrix method, following \cite{Diekmann+H+M-90} and \cite{Diekmann+H-00}, and a force-of-infection analysis based on an estimation of the dominating growth rate (intrinsic rate) of the overall infection pressure in a fully susceptible population. Third, we extend the model to include vaccination policies based on the immunity age, and analyze their efficiency. This enables evaluation of realistic public health policies for booster vaccination.
In addition, births and deaths are included in the model in order to capture potential long-term demographic changes. 

The remainder of this paper is organized as follows. Section 2 presents the mathematical formulation of the dual-age structured model in full detail. Section 3 shows the derivation of solution representations and outlines the numerical solution method. Section 4 presents the basic reproduction numbers using both the next-generation matrix and the force-of-infection approaches. Section 5 extends the framework to include vaccination and demographic factors.
Section 6 presents numerical results exploring model behavior and vaccination strategies under various parameter scenarios.

\section{A dual-age structured epidemiological model} \label{SSEModel}

Our approach tracks all individuals by their immunity age (time since the last recovery from infection) $\tau\in[0,\infty)$  and, additionally for infected individuals, by their infection age (time since infection onset) $\theta\in [0,\infty)$ . 
When individuals become infected, their infection age starts at $\theta=0$ while their immunity 
age $\tau$ remains fixed at the value it had at the time of infection. When individuals recover, 
their immunity age resets to $\tau=0$ and the infection age becomes irrelevant. 

All time variables use a common unit of time, taken as days in the subsequent analysis. 
The age variables $\tau$ and $\theta$ advance at the same rate as calendar time $t$:
\begin{equation}
       \frac{\dd \tau}{\dd t} = \frac{\dd \theta}{\dd t} = 1,
\end{equation}
where $\tau$ advances for susceptible individuals and $\theta$ advances for infected individuals. 

Individuals who have never been infected have $\tau=\infty$ and are considered fully susceptible. In the basic model, we assume that individuals with large $\tau$ become fully susceptible in the limit $\tau\to\infty$. This means we neglect possible differences between fully susceptible individuals and any baseline immunity that might persist after infection for some diseases. Note however, that in Section \ref{SSEModel1} we discuss a model extension that uses a separate class of never-infected individuals and therefore allows baseline immunity to be modeled explicitly.


The pair $(\theta, \tau)$ of age indices implicitly describes the immunity/infection status of individuals in the population. In consequence, all parameter functions like recovery rate or mortality rate depend on the age indices and, implicitly, on the related immunity status. This is quite different from ODE-based SIRS or SEIRS models (see e.g. \cite{KHAN2024657} and \cite{NISHIMURA2023113426}), which use an additional compartment of recovered individuals and a constant transition rate between recovered and susceptible classes to model waning immunity. In our PDE model, the dichotomy between recovered and susceptible individuals is replaced by a continuum of individuals with varying immunity levels, and transition rates depend on the actual immunity status.  


To model population dynamics, we let $S(t,\tau)$ denote the size of the susceptible population at current time $t$ that recovered from the previous infection at time $t-\tau$. Moreover, we let $I(t,\theta,\tau)$ denote the size of the infected population at current time $t$ that was infected at time $t-\theta$ with immunity age $\tau$ at the time of infection. Note that in the current study, we are interested mainly in the short-time aspects (small number of waves) and assume that without any disease the population stays constant (the natural dead are compensated by births). Later, in Section \ref{SSEModel1} we will discuss how to extend the framework in order to model long-term demography in more detail. 

The infectiousness of an infected individual is assumed to be proportional to its pathogen load. The proportionality factor accounts for the infectiousness of the pathogen relative to the size of the socially active population. In the following, $v(\theta,\tau)$ represents the pathogen load, already scaled by the proportionality factor. The dependence on both age variables captures how infectiousness varies during the course of infection and depending on the individual's immunity history.

The other exogenous parameters and functions that appear in the model are:

\bino
$\sigma(\tau)$ -- susceptibility  per contact of a susceptible individual of immunity age $\tau$,\\
$c_0$ -- contact rate of susceptible individuals,\\ 
$\mu(\theta,\tau)$ -- mortality rate of infected individuals,\\
$\rho(\theta,\tau)$ -- recovery rate of infected individuals,\\
$c(\theta,\tau)$ --  contact rate of infected individuals participating in possible transmission.\\

All these parameters and functions take nonnegative values and are assumed to be continuous. If a maximum duration $\bar{\tau}$ of pathogen-specific immunity is assumed, then $\sigma(\tau) = \sigma(\bar{\tau})$ for $\tau \geq \bar{\tau}$, with analogous conditions for $\rho$ and $\mu$. While some of the functions may also depend on time $t$ (e.g. due to seasonality), for the sake of simplicity such dependencies are not included in the following. 

Below, we use the term ``environment" in a restrictive sense, meaning the space in which contacts between individuals take place. The total infectiousness of the environment that susceptible individuals encounter is modeled by
\be \label{EE}  
       E(t) =  \int_0 ^{\infty}\!\!\! \int_0 ^{\infty} v(\theta,\tau) c(\theta,\tau) I(t,\theta,\tau)
            \dd \theta \dd \tau.
\ee
Here, E(t) represents the aggregate infectious potential from all infected individuals at time t, weighted by their pathogen loads and contact rates. In the following, we call E(t) the \emph{infection pressure}.

Alternatively, if population size variations are important, one can normalize the above expression by the active population size as follows:
\bd
         E(t) = \frac{\int_0^{\infty}\!\!\! \int_0 ^{\infty} v(\theta,\tau) c(\theta,\tau) I(t,\theta,\tau)
            \dd \theta \dd \tau}{c_0 \int_0^\infty S(t,\tau) \dd \tau +
              \int_0 ^{\infty}\!\!\! \int_0^{\infty} c(\theta,\tau) I(t,\theta,\tau) \dd \theta \dd \tau}.
\ed
However, in the sequel we follow, i.e., \cite{Ghosh+Volpert-23}, using the simpler expression \reff{EE}.

Building on these considerations, the dynamics of compartments $S$ and $I$, stratified by immunity 
age and infection age, can be described by a system of age-structured partial differential equations. 
Each equation describes the transport of population densities through their respective aging coordinates 
at unit velocity on the left-hand sides, combined with the in-/out-flow due to epidemiological factors (infection, recovery, 
death) on the right-hand side: 
\bea 
  \label{ES}         \frac{\partial S(t,\tau)}{\partial t} + \frac{\partial S(t,\tau)}{\partial \tau}
            &=& - \sigma(\tau) c_0 E(t) S(t,\tau),\\
  \label{EI}  \frac{\partial I(t,\theta,\tau)}{\partial t} + \frac{\partial I(t,\theta,\tau)}{\partial \theta}
            &=& - (\mu(\theta,\tau) + \rho(\theta,\tau)) I(t,\theta,\tau),
\eea
where $t, \theta, \tau \in [0,\infty)$. Note that $E$ is a nonlocal term depending on the whole distribution
of the infected individuals $I$ over immunity and infection ages at time $t$. The system therefore is non-linear.

The initial conditions are
\be \label{ES0}
             S(0,\tau) = S^0(\tau),  \quad I(0,\theta,\tau) = I^0(\theta,\tau), 
                                 \quad \theta,\tau \in [0,\infty),
\ee
where $I^0, S^0$ represent the initial distribution of infected and susceptible individuals with respect to age variables.

Without loss of generality, we assume in the following that the initial population is normalized to one:
\bd
    \int_0^{\infty} S^0(\tau) \dd \tau +
          \int_0 ^{\infty}\!\!\! \int_0 ^{\infty} I^0(\theta,\tau) \dd \theta \dd \tau =1.
\ed

\begin{figure}
    \centering
    \includegraphics[width=0.75\linewidth]{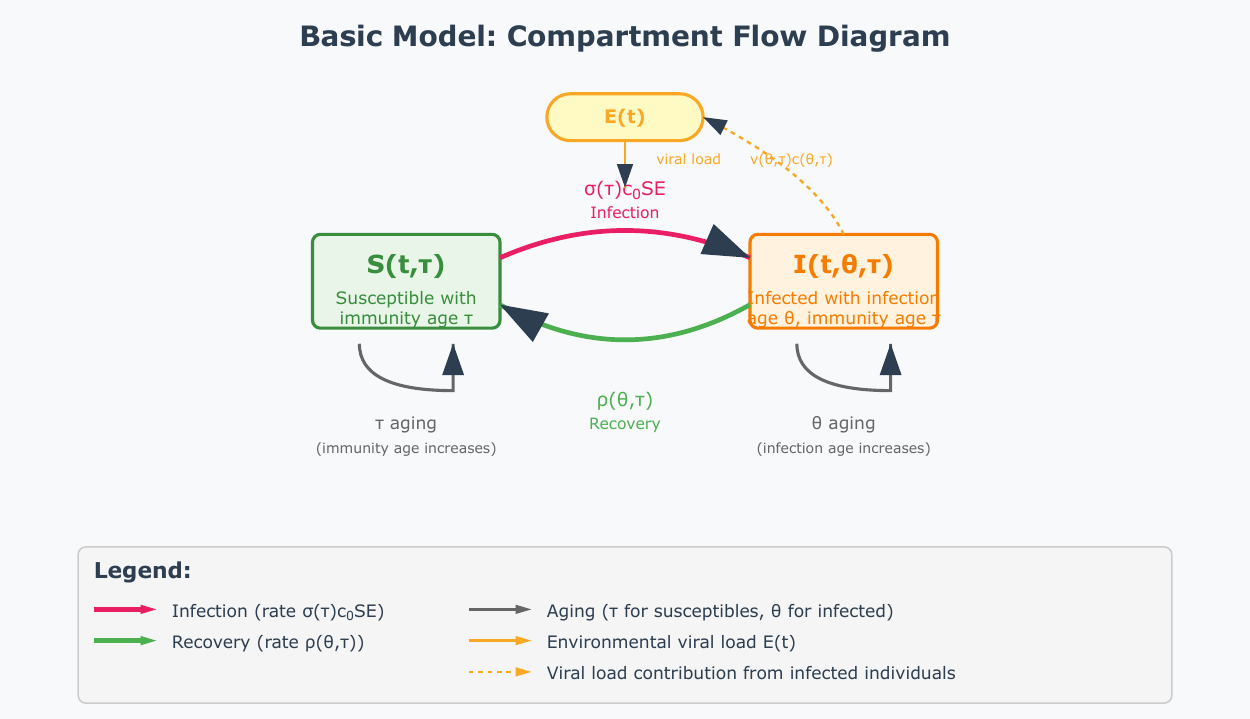}
    \caption{Structure of the basic dual-age epidemiological model. Note that immunity age $\tau$ increases only in the susceptible class and infection age $\theta$ increases only in the infected class. Infected individuals with any combination $(\theta,\tau)$ of ages have new immunity age $\tau = 0$ after recovery.}
    \label{fig:basic}
\end{figure}

Finally, the aging processes need boundary conditions at $\tau=0$ and $\theta=0$: all newly 
recovered individuals enter the susceptible population at $\tau=0$, while newly infected individuals 
enter the infected population with $\theta=0$. The recovery flux B(t) quantifies the total rate of recovery 
across all infected individuals, representing the biological process by which the infected population 
feeds back into the susceptible population. Note that newly infected individuals keep the information 
about their immunity age at infection, whereas newly recovered individuals lose the information about 
their infection age, and all of them start at immunity age 0. These observations result in the 
following nonlocal boundary conditions:
\bea
    \label{ESt0}   S(t,0) &=& B(t) :=  \int_0^{\infty}\!\!\! \int_0^{\infty} \rho(\theta,\tau) I(t,\theta,\tau) 
                  \dd \theta \dd \tau,\\
    \label{EIt0}        I(t,0,\tau) &=& \sigma(\tau) c_0 E(t) S(t,\tau).
\eea
Figure \ref{fig:basic} summarizes the essential model structure. 

\bino

The susceptibility parameter $\sigma(\tau)$, being an increasing function of $\tau$,
describes how the immunity wanes with the time since recovery. The waning can be caused 
by the in-body loss of immunity or by relative loss of immunity due to mutations of the pathogen.
In both cases, often the pathogen-specific immunity practically disappears in a finite time $\bar \tau > 0$. 
This means that the function $\sigma(\tau)$ becomes nearly constant for $\tau > \bar \tau$, with $\sigma(\bar \tau)$ corresponding to baseline immunity.
In general, since $\sigma$ can be assumed bounded, the monotonicity implies that 
$\lim_{\tau \ra +\infty} \sigma(\tau) = \sup_{\tau \geq 0} \sigma(\tau)$. Hence, 
for every $\e > 0$ there is $\bar \tau \geq 0$ such that 
$\sigma(\tau) \in [\sigma(\bar \tau),\sigma(\bar \tau) + \e]$ for all $\tau \geq \bar \tau$.
Then one can reasonably assume that $\sigma$ is constant after immunity age $\bar \tau$.
This simplification is numerically justified by a computation done in Section \ref{SNumer}, where the sensitivity of the solution with respect to $\bar \tau$ is estimated. 

One can avoid the consideration of integrals on infinite horizons---which is particularly important for numerical implementation--- in the following way.
Assuming that $S^0(\tau) = 0$ for $\tau > \bar \tau$, we have $S(t,\tau) = 0$ for every $t\in [0,T]$,
$\tau > t+\bar \tau$.\footnote{%
This assumption is not restrictive. Even if some of the initially susceptible individuals have $\tau > \bar \tau$, in order to avoid a $\delta$-function concentrated at $\bar \tau$ we can redefine $\bar \tau$ as $\bar \tau_1 := 1.01 \bar \tau$ and consider the initial population with $\tau > \bar \tau$
as uniformly distributed in $[\bar \tau, \bar \tau_1]$. This does not change the problem since all data are constant for $\tau > \bar \tau$. The same applies to $I$.} Then one can replace the integration with respect to $\tau$ in 
equations \reff{EE} and \reff{ESt0} with integration on $[0, t+\bar \tau]$. Moreover, if $\bar \theta > 0$ is
the maximal duration of the disease (excluding abnormal cases), say 20 days for COVID-19, then the size of the infected population of larger infection age must be zero. Usually, when modeling finite horizon $\bar\theta$, recovery and/or mortality rates are defined as tending to infinity at $\bar\theta$. However, as an alternative, we may just assume that all still existing infected individuals recover at infection age $\bar\theta$.  
In this case, the implementation of the finite horizons approach only requires replacement of equation \reff{ESt0} with
\bd
      S(t,0) = B(t) := \int_0^{t+\bar \tau}\!\!\! \int_0^{\bar \theta} \rho(\theta,\tau) I(t,\theta,\tau) 
                  \dd \theta \dd \tau + \int_0^{t+\bar \tau} I(t,\bar \theta,\tau) \dd \tau,\\
\ed
and equation \reff{EE} becomes
\bd
      E(t) =  \int_0 ^{t+\bar \tau}\!\!\! \int_0 ^{\bar \theta} v(\theta,\tau) c(\theta,\tau) I(t,\theta,\tau)
            \dd \theta \dd \tau,\\        
\ed

\begin{rem} \label{Rdeads}{\em
     One can easily calculate the number of currently dying individuals as follows:
\bd
  D(t) = \int_0^{t+\bar \tau} \int_0^{\bar \theta} \mu(\theta,\tau) I(t,\theta,\tau) \dd \theta \dd     \tau.            
\ed
}\end{rem}

Further on, we use the shortcuts $\kappa=\mu+\rho$, $v_c= v \, c$ and $\sigma_c=\sigma \,c_0$. 

\section{Representation of the solution and numerical solution} \label{S_Sol}

First, we discuss the meaning, the existence, and the representation of a solution  to the system
\eqref{EE}--\eqref{EIt0}.
Denoting $\eta = (1,1)$ and $\zeta = (1,1,0)$, we define
\bd
        \D_\eta S(t,\tau) := \lim_{\al \ra 0+} \frac{S((t,\tau) + \al \eta) - S(t,\tau)}{\al}, \quad
         \D_\zeta I(t,\theta,\tau) := \lim_{\al \ra 0+}\frac{I((t,\theta,\tau) + \al \zeta) - I(t,\theta,\tau)}{\al}, 
\ed
provided that the limits in these directional derivatives exist. Then the traditional notations on the left-hand sides of equations 
\reff{ES} and \reff{EI} should be understood as $\D_\eta S(t,\tau)$ and $\D_\zeta I(t,\theta,\tau)$,
respectively. A solution of \eqref{EE}--\eqref{EIt0} then is a measurable pair $(S, I)$ for which
the limits in the definitions of directional derivatives exist for almost every $(t,\tau) \in \Omega_S := [0,\infty) \times [0,\infty)$, resp. 
$(t,\theta,\tau) \in \Omega_I :=  [0,\infty) \times [0,\infty)\times [0,\infty)$, the integrals in 
\reff{EE} and \reff{ESt0} are finite, and all the equations are satisfied almost everywhere.
In particular, a solution does not need to be differentiable neither in $t$ nor in 
$\tau$ or $\theta$ and can even be discontinuous
(for example if the initial and the boundary data are not consistent). 

Theorem 1 in \cite{VV:JMAA_08} implies
that a unique solution exists (the required linear growth condition in this theorem is formally not satisfied in our case, but it is not needed because the function $E$ is bounded). 
Moreover, Theorem 2 in the same paper
claims that the problem is well-posed on any finite time-interval.
More precisely, $L_\infty$-disturbances of any of the parameters lead to at most
proportional $L_\infty$-change of the solution. Similarly, 
$L_1$-disturbances result in $L_1$-change of the solution.\footnote{%
We mention that the parameters should be considered as represented by the function $u$ in \cite{VV:JMAA_08}.}

System \eqref{EE}--\eqref{EIt0} is nonlinear due to the dependence of $E$ and $B$ on $I$, but thanks to the linearity directly with respect 
$S$ and $I$ one can obtain convenient representations of the solution as long as $E(t)$ and $B(t)$ are considered as given. For that, we notice that 
for any scalar functions $g$ (of one scalar variable) and $h$ (of two scalar variables), the 
following obvious identities hold:
\bd
       \D_\eta g(\tau-t) = 0, \qquad  \D_\zeta h(\theta-t,\tau) = 0. 
\ed   
Thanks to these identities, a direct check verifies that 
\bea
    \label{ESolS}   S(t,\tau) &=& z_S(t-\tau) 
       e^{-\int _{\max\{0, t-\tau\}}^t E(\beta) \sigma_c (\beta+\tau-t)\ddd \beta},\\
  \label{ESolI}      I(t,\theta,\tau) &=& z_I(t-\theta,\tau) 
        e^{-\int _{\max\{0, t-\theta\}}^t\kappa (\beta+\theta-t,\tau)\ddd \beta},
\eea
where
\bd
        z_S(\al)= \left\{ \begin{array}{cl}
                     S^0(-\al) & \mbox{ if } \; \al \leq 0,\\
                     B(\al) &   \mbox{ if } \; \al > 0,      
                     \end{array} \right. \qquad
         z_I(\al,\tau) = \left\{ \begin{array}{cl}
                     I^0(-\al,\tau) & \mbox{ if } \; \al \leq 0,\\
                     c_0 \sigma(\tau)S(\al,\tau) E(\al)  &   \mbox{ if } \; \al > 0,      
                     \end{array} \right.                 
\ed
solve \reff{ES}--\reff{EI} on $\Omega_S$, resp. $\Omega_I$, and also the initial and boundary conditions \reff{ES0}--\reff{EIt0}
are satisfied. 

In particular, in the next section we shall use the representation of $I$:
\bea\label{ISol1}
       I(t,\theta,\tau) = I^0(\theta-t,\tau) e^{-\int _0^t\kappa (\beta+\theta-t,\tau)\ddd \beta} 
                 \qquad \mbox{ if }   \;\;  \theta \geq t,
\eea
and
\bea\label{ISol2}
    \label{solI1a}I(t,\theta,\tau) &=& \sigma_c (\tau ) E(t-\theta ) S(t-\theta ,\tau ) e^{-\int _{t-\theta }^t\kappa (\beta+\theta-t,\tau )\dd \beta}\nonumber\\
    &=& \sigma_c (\tau ) E(t-\theta ) S(t-\theta ,\tau ) e^{-\int _{0 }^{\theta}\kappa (\beta,\tau )\dd \beta} \qquad\text{ if } \;\; \theta < t.
\eea
Formula \eqref{ISol1} represents individuals who were initially infected and remain infected at time $t$, while \eqref{ISol2} represents individuals who became infected after the initial time. 
Note that the function $\theta\mapsto e^{-\int _{0 }^{\theta}\kappa (\beta,\tau )\dd \beta}$, the exponential term in \eqref{ISol2}, is the ``survival function" for staying infected. 

Finally, plugging \eqref{ESolS}, \eqref{ESolI} into \eqref{EE} and \eqref{ESt0} results in a system of two nonlinear integral equations for $E(t)$ and $B(t)$. Substituting the resulting solution back into \eqref{ESolS} and \eqref{ESolI} gives the solution of the system \eqref{EE}--\eqref{EIt0}. 

{\bf Notes about the numerical method.} In the current study we use a direct approach for solving the system, namely discretization along the characteristic lines associated with 
equations \reff{ES}--\reff{EI}. 

Denote by 
\bda
    && \Gamma_S^0 = \{(0,\tau_0) \sth \tau_0 \in [0,\bar \tau]\}, \qquad
     \Gamma_S^b = \{(t_0,0) \sth t_0 \in [0,T]\}, \\
    && \Gamma_I^0 = \{(0,\theta_0,\tau) \sth \theta_0 \in [0,\bar \theta], \; \tau \in[0,\bar \tau]\}, \qquad
     \Gamma_I^b = \{(t_0,0,\tau) \sth t_0 \in [0,T], \; \tau \in [0,t_0+\bar \tau]\},
\eda
the sets where the initial/boundary conditions for $S$ and $I$, respectively, are given.
The characteristic lines starting from the initial/boundary points that fill in the domains 
$\Omega_S$ and $\Omega_I$, have the form $\{a + \al \eta \sth \al \geq 0\}$, 
$\{b + \beta \zeta \sth \beta \geq 0\}$, where $a := (t_0,\tau_0) \in \Gamma_S^0 \cup \Gamma_S^b$,
$b := (t_0,\theta_0,\tau) \in \Gamma_I^0 \cup \Gamma_{I}^b$. On every characteristic line, the solutions 
$\al \mt S[a](\al)$ and $\beta \mt I[b](\beta)$ satisfy the ordinary differential equations\footnote{%
\label{FnSolSI} These ODEs are actually used to derive the representation \reff{ESolS}--\reff{ESolI}
up to the initial/boundary multipliers $z_S$ and $z_I$.}
\bda
     \frac{\dd}{\dd \al} S[a](\al) &=& - \sigma_c(\tau_0 + \al)E(t_0+\al) S[a](\al), \\
      \frac{\dd}{\dd \beta} I[b](\beta) &=& -\kappa(\theta_0+\beta, \tau) I[b](\beta).
\eda

We discretize these ODEs using the Heun scheme -- an explicit second order Runge-Kutta scheme with second
order convergence. A computational challenge arises because the boundary conditions are endogenous and non-local -- depending on the integrals of $S$ and $I$ with respect to $\tau$ or
$\theta$ and $\tau$, respectively. Therefore, at every time-step of the scheme we compute $S$ and $I$ at the
subsequent time with a first-order accuracy, evaluate the needed integrals, then recalculate $S$ and $I$ in order to preserve the second order convergence of the overall discretization scheme.

The method was implemented in Matlab\textsuperscript{\texttrademark}, version R2019b (64bit). In the experiments described in Section \ref{SNumer}, the solution process for one instance with satisfactory accuracy takes between 10 and 60 seconds at a commercially available notebook.

%

\section{Basic reproduction numbers} \label{SBRN}
In the following we derive basic reproduction numbers, based on two classical approaches: the next-generation approach and the derivation of a basic reproduction number, together with an intrinsic growth rate from the related renewal equation under the assumption of a fully susceptible population. Both approaches are based on the infection pressure \eqref{EE}. The first tracks secondary infections directly caused by an initial infected cohort---yielding a worst-case measure for introduction risk. The second analyzes the asymptotic growth rate after the initial generation has passed through---providing the intrinsic growth rate in a fully susceptible population. 

\paragraph{The next generation approach.} Let a ``small" portion of infected individuals, $I^0(\theta,\tau)$, $\theta,\tau \in [0,\infty)$,
with $\int_0^{\infty}\!\!\! \int_0 ^{\infty} I^0(\theta,\tau) \dd \theta \dd \tau =: \hat I^0$,
be inserted in a susceptible population
$S^0(\tau)$ with $\int_0^{\infty} S^0(\tau) \dd \tau = 1$.   
Following the  \emph{next generation} methodology (see \cite{Diekmann+H-00})
for evaluation of the Basic Reproduction Number (BRN),
we assume that at any time $t$, the infected population acts on the susceptible population 
as there were no infection before $t$. However, the susceptible population changes with time 
even without infection, namely, according to \eqref{ES},
\be \label{EH541}
 S(t,\tau) = \left\{\begin{array}{cl}
   S^0(\tau-t) & \mbox{ if } \, \tau \in [t,\infty),\\
                  0 &  \mbox{ if } \, \tau \in [0,t).
    \end{array} \right.             
\ee
 
Next, we evaluate the direct contribution, $E_0(t)$, of the initially infected population $I^0$
to the infection pressure at time $t$. 
An infected individual with some $(\theta,\tau)$ at time $t$ 
belongs to the initially infected population if and only if $\theta \geq t$, therefore
\bd
         E_0(t) = \int_0^{\infty}\!\!\! \int_t ^{\infty} v_c(\theta,\tau) I(t,\theta,\tau) 
            \dd \theta \dd\tau,
\ed
where $I$ is the solution of  \eqref{EI} with initial condition $I^0$.
Using \eqref{ISol1} 
we obtain the expression
\be \label{EV0}
           E_0(t) := \int_0^{\infty}\!\!\! \int_t ^{\infty} v_c(\theta,\tau) 
                    e^{-\int_0^t \kappa(\theta-(t-s),\tau) \ddd s} \,I^0(\theta-t,\tau)  \dd \theta \dd\tau.
\ee
Hence, applying \eqref{EIt0}, the new infections at time $t$ caused by the originally 
infected population can be expressed as
\bd
          E_0(t) \int_0^{\infty}  \sigma_c(\tau) S(t,\tau) \dd \tau  = 
          E_0(t) \int_t^{\infty} \sigma_c(\tau) S^0(\tau-t) \dd \tau,
\ed 
where in the last equality we utilize \reff{EH541}.
Thus we represent the BRN 
(new infections directly caused by a unit of the original infected populations) as
\bd
       \mathcal{R}_0[S^0(\cdot),I^0(\cdot)] = \int_0^{\infty} \!\!\! \int_t ^{\infty} 
         \sigma_c(\tau) S^0(\tau-t) \dd \tau \, \frac{E_0(t)}{\hat I^0}\dd t,
\ed
where $E_0$ is given by \eqref{EV0}. 

The calculation of the BRN $\mathcal{R}_0[S^0(\cdot),I^0(\cdot)]$ 
requires exact information about the distribution of the susceptible and infected populations.
Following e.g. \cite{Diekmann+H+M-90}, we may define BRN by the model parameters only, 
considering the worst case scenario for the distribution of $S^0$ and $I^0$.  Changing the 
variable $\tau$ and the order of integration one can represent
\be \label{ERN_SI}
       \mathcal{R}_0[S^0(\cdot),I^0(\cdot)] = \int_0^{\infty}  \!\!\!\int_0^{\infty} 
         \sigma_c(t+\al) \frac{E_0(t)}{\hat I^0} \dd t \,  S^0(\al) \dd \al.
\ee
The worst case for $S^0$ is that all initial population have no immunity (that is, to assume
that $\sigma$ is monotone increasing, which corresponds to the fact that immunity is waning). Then, in absence of information about the distribution of $S^0$, one may define the BRN as 
\bea
   \nonumber    \mathcal{R}_0[I^0(\cdot)] &=&  \sigma_c(\bar \tau)  \int_0 ^{\infty} 
          \frac{E_0(t)}{\hat I^0} \dd t \\
 \nonumber         &=& \sigma_c(\bar \tau) 
     \int_0 ^{\infty} \int_0^{\infty}\!\!\! \int_t ^{\infty} v_c(\theta,\tau) 
                    e^{-\int_0^t \kappa(\theta-(t-s),\tau) \ddd s} \,\frac{I^0(\theta-t,\tau)}{\hat I^0} \dd \theta \dd\tau \dd t\\
    \label{EH65}   &=&  \sigma_c(\bar \tau) \int_0 ^{\infty} \!\!\!
        \int_0^{\infty}\!\!\! \int_0^{\infty} v_c(t+\al,\tau)
                    e^{-\int_0^t \kappa(\al+s,\tau) \ddd s} \dd t \, \, 
            \frac{I^0(\al,\tau)}{\hat I^0} \dd \alpha \dd\tau.
\eea

Now, consider also the worst case with respect to the distribution of $I^0$, which gives                                                             
the BRN without knowledge of the initial distributions of susceptible and
infected populations:
\bea \nonumber
       \mathcal{R}_0 &=& \sigma_c(\bar \tau) \max_{\al,\tau \geq 0} \int_0^{\infty} 
        v_c(t+\al,\tau) e^{-\int_0^t \kappa(\al+s,\tau) \ddd s} \dd t \\
     \label{ER0}   &=& \sigma_c(\bar \tau) \max_{\al,\tau \geq 0} \int_\al^{\infty}
     v_c(\theta,\tau) e^{-\int_\al^\theta \kappa(\zeta,\tau) \ddd \zeta} \dd \theta.                  
\eea

If it is known in advance that all initially infected individuals (can be a single person) have
$\theta = 0$ (that is, the infection is internally generated), then one should set $I^0(\al,\tau) = \delta_0(\al)I^0(\tau)$ in \reff{EH65}.
(Here $\delta_0(\cdot)$ is the Dirac $\delta$-function concentrated at zero.)
Since in this case it is natural to assume that the infected 
individuals have no adaptive immunity   (that is, $I^0(\tau) = \delta_0(\tau-\bar \tau) \hat I^0$), formula \reff{ER0} reduces to
\be \label{ER02}
       \hat{\mathcal{R}}_0 =  \sigma_c(\bar \tau) \int_0^{\infty}
     v_c(\theta,\bar \tau) e^{-\int_0^\theta \kappa(\zeta,\bar \tau) \ddd \zeta} \dd \theta,    
     \ee
where $\bar \tau$ can also be $+\infty$.

If the numbers $\bar \theta$ and $\bar \tau$ are finite, then the above infinite-horizon integrals become finite: the triple integral in \reff{EH65} becomes $\int_0^{\bar \tau} \int_0^{\bar \theta}\int_0^{\bar \theta-\theta}$, similarly in \reff{ER0} and \reff{ER02}.

If all data are independent of $\theta$ and $\tau$, in which case the model \eqref{EE}--\eqref{EIt0}
reduces to the standard $S$-$I$ model, the expression in \eqref{ER0} and hence also in \eqref{ER02} becomes 
$\frac{c_0\sigma v c}{\mu+\rho}$, which coincides with the usual expression 
of the BRN in our notations.

\bino
The derivation of the above formulae
related to the reproduction rate of the disease gives advice for their practical utilization. If information is available about the distributions of the initial data, then the value $\mathcal{R}_0[S^0(\cdot),I^0(\cdot)]$ 
gives precise information on whether the disease is declining or expanding.
If such information is available only for $I^0$, then $\mathcal{R}_0[I^0(\cdot)]$ can be used
as an upper estimation for $\mathcal{R}_0[S^0(\cdot),I^0(\cdot)]$.
If information for the distribution of $I^0$ is also not available, then one can use either \reff{ER0} or \reff{ER02}
depending on whether the disease is 
``imported" or is internally generated.
In both cases $\mathcal{R}_0 < 1$, respectively $\hat{\mathcal{R}}_0 < 1$ guarantees that the disease shrinks, however, values larger than one do not imply that the disease initially expands.

\paragraph{The intrinsic growth rate approach:} We analyze now the infection pressure after removal of all initially infected individuals. There are relations to the methodology in \cite{Diekmann+H+M-90}, \cite{HeesterbeekDietz-96}, but also differences in the model specification and the derivation starts with an integral equation for the infection pressure, derived from the PDE-system, instead of the next generation operator.  Moreover, we make the assumption that infections occur within a fully susceptible population throughout. In this setting, all individuals have $\tau=\infty$, which is achieved by $S(t,\tau)=\delta_{\infty}(\tau)$ with $\delta_\infty$ denoting the limiting Dirac-delta distribution at infinity.  

We consider $t\geq \bar{\theta}$, where $\bar{\theta}$ denotes the maximal infection age. At $t=\bar{\theta}$ all infected individuals of the first generation are either recovered or dead and we analyze the further dynamics of the epidemics within a (again) fully susceptible population. This is quite different from the next generation approach, which focuses exactly on the first generation period.

Only the solution \eqref{ISol2} is relevant for $t\geq \bar{\theta}$. Plugging it into \eqref{EE}, and using the assumption of a fully susceptible population leads to the homogeneous renewal equation
\bea \label{renewal}
    \hat{E}(t)&=&\int _0^t \int _0^{\infty}\delta_{\infty}(\tau)v_c(\theta, \tau)\sigma_c (\tau ) \hat{E}(t-\theta ) 
     e^{ -\int _{0 }^{\theta}\kappa (\beta ,\tau)\ddd \beta } \dd \tau \dd \theta\nonumber\\
     &=&\sigma_c^\infty \int _0^t v_c^\infty(\theta)\hat{E}(t-\theta ) 
     e^{ -\int _{0 }^{\theta}\kappa^\infty (\beta)\ddd \beta } \dd \theta\\
     &=&\int _0^t \hat{E}(t-\theta ) h(\theta) \dd \theta,\nonumber
\eea
where $\hat{E}$ denotes the infection pressure in a fully susceptible population for $t\geq\bar{\theta}$ and 
$$
h(\theta)=\sigma_c^\infty v_c^\infty(\theta)
e^{-\int _{0 }^{\theta}\kappa^\infty (\beta)\ddd \beta}.
$$ 
Here, we also used the properties of the Dirac-delta and introduced the notation $v_c^{\infty}, \kappa^{\infty}, \sigma_c^{\infty}$ to indicate the limits $\tau\to\infty$ of the parameter functions. The renewal equation \eqref{renewal} is derived from a reduced system, where the 
susceptible dynamics are frozen at the disease-free equilibrium state. This linear equation captures 
the infection dynamics when susceptible depletion is negligible.

Applying now the Laplace transform to both sides of the equation and accounting for the convolution form of the integral leads to the equation
\bda
    \mathcal{L}\{\hat{E}\}(r)=\mathcal{L}\{\hat{E}\}(r)\mathcal{L}\{h\}(r).
\eda
This shows that nontrivial solutions to the renewal equation exist only if 
\bea\label{EuLo}
    \label{rCond2}1=\mathcal{L}\{h\}(r)=\sigma_c^{\infty}\int _0^\infty e^{-r\theta}  v_c^{\infty}(\theta) 
    e^{-\int_0^{\theta } \kappa^{\infty} (u) \, \ddd u} \dd \theta,
\eea 
which is the Euler-Lotka equation in our setting. 

If $v_c^\infty$ is bounded and $\int_0^{\infty } \kappa^{\infty} (u) \, \ddd u = \infty$, then there exist solutions. For the reasonable 
case of $\sigma_c^{\infty}, v_c^{\infty}(\theta)\geq 0$, the right-hand side of \eqref{EuLo} 
is a continuously decreasing function of $r$ with limit $+\infty$ as $r \to -\infty$ and limit $0$ 
as $r \to +\infty$. Therefore, there exists a unique real solution of \eqref{EuLo}, which we denote by $r^*$.

Generally, there will be additional complex characteristic exponents $r$ satisfying \reff{EuLo}. Any linear combination of $E_0 e^{r^*t}$ and $E_j e^{r_j^*t}$ with additional solutions $r_j$ to \eqref{EuLo} is a solution of the renewal equation. In principle the correct linear combination has to be chosen such that $\hat{E}(t)$ for $t \geq \bar{\theta}$ fits to $\hat{E}(t)$ for $t \leq \bar{\theta}$, which in turn is determined by the initially infected population $I_0(\theta,\tau)$. However, $r^*$ is the solution with the largest real
part\footnote{
Let $\lambda = \alpha + i\omega$ with $\omega \neq 0$ be a complex solution 
to \eqref{EuLo}. By the triangle inequality, and using positivity of $h$,
$$|\mathcal{L}\{h\}(\lambda)| = \left|\int_0^\infty e^{-\alpha\theta} e^{-i\omega\theta} 
h(\theta) d\theta\right| < \int_0^\infty e^{-\alpha\theta} h(\theta) d\theta 
= \mathcal{L}\{h\}(\alpha)$$
for $\omega \neq 0$. Thus $\mathcal{L}\{h\}(\alpha) > 1$. 
Since $\mathcal{L}\{h\}$ is strictly decreasing and $\mathcal{L}\{h\}(r^*) = 1$, 
we conclude $\alpha < r^*$.}, and hence $e^{r^*t}$ is the dominating behavior. One may interpret $r^*$, therefore, as the intrinsic growth rate, in a fully susceptible population, see \cite{Sharpe+1911}. 

Based on \eqref{EuLo}, the same properties that guarantee a unique real  solution $r^*$ also imply that 

\bea\label{R01}
    \sigma_c^{\infty}\int _0^\infty  v_c^{\infty}(\theta) e^{-\int_0^{\theta } \kappa^{\infty} (u)\, 
    \, \ddd u} \dd \theta=\hat{\mathcal{R}}_0
\eea
indicates the sign of $r^*$: if $\hat{\mathcal{R}}_0<1$ then $r^*<0$ and the infection pressure will shrink after the first generation has been removed, even in a fully susceptible population. When $\hat{\mathcal{R}}_0>1$ then $r^*>0$, which indicates growth of the infection pressure after the first generation has been removed, as long as the susceptible population stays large enough. 

The integral is well defined if  $v_c^{\infty}(\cdot)$ is bounded and
$\kappa^{\infty}(u) \geq \kappa_0 > 0$ is bounded away from zero, since then the exponential term provides exponential decay. Moreover, $\mathcal{R}_0$ can also be interpreted as the expected number (via the survival function) of secondary infections over the infectious period, caused by one infected individual with $\tau=\infty$ in a fully susceptible population.

For the infinite horizon case $\bar{\theta}=\infty$, one may add the following considerations: Choose a large $\bar{\theta}$. Then use the truncated survival function $\mathcal{S}_{\bar{\theta}}(\theta) = e^{-\int_0^\theta \kappa^\infty(\beta) \, \ddd \beta} \cdot \mathds{1}_{\theta \leq \bar{\theta}}$ for calculating $r^*$ and $\hat{\mathcal{R}}_0$ with integral boundary $\bar{\theta}$ instead of $\infty$. This results in values $r^*(\bar \theta)$ and $\hat{\mathcal{R}}_0(\bar{\theta})$. Clearly, $\hat{\mathcal{R}}_0(\bar{\theta}) = \sigma_c^\infty \int_0^{\bar{\theta}} v_c^\infty(\theta) e^{-\int_0^\theta \kappa^\infty(u) du} d\theta$ converges to $\mathcal{R}_0$ as $\bar{\theta} \to \infty$. Moreover, $r^*(\bar{\theta})$ is decreasing in $\bar{\theta}$ with the infimum $\bar{\theta}$, which implies $r^*(\bar{\theta})\to r^*$ if $\bar{\theta} \to \infty$.

Note that $\hat{\mathcal{R}_0}$ was derived in \eqref{R01} from a different starting point and under different assumptions than in \eqref{ER02}: the next generation approach was based on \eqref{ISol1} which describes infection from the first generation of infected. On the other hand, the intrinsic growth rate calculations stem from \eqref{ISol2}, and hence are related to the effects after the first generation of infected individuals has vanished. Moreover, \eqref{ER02} was derived under the specific assumption $\theta=0$ for all individuals in the first generation of infected, whereas \eqref{ER0} does not assume anything about the first (or any) generation of infected but analyzes the effect of the disease in a permanently naive population. The intrinsic growth rate approach therefore adds another interpretation of $\hat{\mathcal{R}}_0$: it is also reasonable in situations where a large part of the population remains fully susceptible (even after several initial generations of infected individuals. Moreover, the approach provides the exponential growth rate $r^*$ in a naive population.

\section{Model extensions} \label{SExtent}

The basic dual-age structured model \eqref{EE}--\eqref{EIt0} can be extended in several directions.
In this section we present two such extensions: (i) incorporation of vaccination policy regarding the immunity age of individuals; (ii) 
consideration of demographic factors
(births and natural deaths) that are relevant for description of the long run demographic dynamics in presence of endemic disease. Later on, the vaccination model will be compared with the basic model \eqref{EE}--\eqref{EIt0}



\subsection{A model with vaccination} \label{SSVacMod}

Vaccination represents an important public health intervention and can be easily incorporated into the dual-age framework. A key advantage of such an extension is the ability to naturally represent immunity-age-based vaccination targeting, reflecting real-world vaccination programs that already prioritize individuals based on time since their last vaccination or infection (booster campaigns). 

Therefore, based on \eqref{EE}--\eqref{EIt0}, we model vaccination as an immunity-boosting process that resets an individual's immunity 
age to zero, similar to recovery from infection. To this end, we introduce a vaccination rate $u(t,\tau)$, representing the vaccination effort per capita at time $t$ directed 
towards susceptible individuals with immunity age $\tau$. 

Usually, only non-infected individuals are vaccinated. The vaccination extension therefore requires modifications of equations \eqref{ES} and \eqref{ES0} as follows:
\bda 
  \label{ESv}         \frac{\partial S(t,\tau)}{\partial t} + \frac{\partial S(t,\tau)}{\partial \tau}
            &=& - \sigma(\tau) c_0 E(t) S(t,\tau) - u(t,\tau) S(t,\tau),\\
  \label{ES0v} S(t,0) &=&  \int_0^{\infty}\!\!\! \int_0^{\infty} \rho(\theta,\tau) I(t,\theta,\tau) 
                  \dd \theta \dd \tau + \int_0^\infty u(t,\tau) S(t,\tau) \dd \tau.
\eda
The first equation models individuals with immunity age $\tau$ leaving the $S(t,\tau)$ compartment through both infection and vaccination.  
The second equation captures that newly vaccinated individuals join those recovering from infection 
in entering the $S(t,0)$ compartment with renewed immunity. Notice that the basic reproduction number $\mathcal{R}_0[S^0(\cdot),I^0(\cdot)]$ given by formula \reff{ERN_SI} changes if vaccination is applied (reflected in the immunity-age structure of $S^0$). However, both ``worst case" basic reproduction numbers $\mathcal{R}_0[I^0(\cdot)]$ and $\mathcal{R}_0$ are not influenced by vaccination, because it affects 
only~$S$. 

Booster vaccination strategies 
are vaccination strategies, where individuals become eligible for vaccination only after their immunity has waned sufficiently over a specified 
period $\tau^*$. In our framework it may be modeled by
\begin{equation}
              u(t,\tau) = \left\{\begin{array}{cl}
                          u^* & \text{ if } \tau \geq \tau^*, \\
                          0  & \text{ if } \tau \in [0,\tau^*),
                              \end{array} \right.
\end{equation}
where $u^*$ represents the vaccination rate and $\tau^*$ is the immunity age threshold above 
which vaccination is recommended.  

In the above extension, which we will use in our numerical simulations, we assume that vaccination has the same effect on immunity as 
infection. This is not realistic for many known epidemic diseases. Therefore, in the next lines 
we present a further extension, taking into account that vaccination improves immunity, but not
as good as infection, in particular, after vaccination the individuals do not return to immunity age
$\tau = 0$. For this reason, we introduce a function $\ph:[0,\infty) \to [0, \infty)$ representing
the new immunity age, $\ph(\tau)$, after vaccination of an individual of immunity age $\tau$.
This function has naturally the properties $\ph(0) = 0$, $\ph(\tau) \leq \tau$, 
and $\ph(\tau_1) \leq \ph(\tau_2)$ whenever $\tau_1 < \tau_2$. Let us assume, additionally,
that the function $\ph$ is continuously differentiable and $\ph'(\tau) > 0$. Then a simple
conservation of mass argument leads to the equation 
\be \label{Evacc2}
    \frac{\partial S(t,\tau)}{\partial t} + \frac{\partial S(t,\tau)}{\partial \tau}
            = - \sigma(\tau) c_0 E(t) S(t,\tau) - u(t,\tau) S(t,\tau) + 
       u(t,\ph^{-1}(\tau)) S(t,\ph^{-1}(\tau)) \frac{1}{\ph'(\tau)}. 
\ee
The other equations in the model \eqref{EE}--\eqref{EIt0} remain the same.
However, $\ph$ is an additional functional parameter 
that has to be known. One can simplify the situation, by assuming that $\ph(\tau) = \tau_0$
for any $\tau \geq \tau_0$, where $\tau_0 \geq 0$ has a clear meaning: the vaccination 
creates immunity of age $\tau_0$ no matter what is the current immunity of the individuals with $\tau > \tau_0$. In the case $\tau_0 = 0$ we introduced in \reff{ES0v} a modification 
(a jump) of the boundary condition. If $\tau_0 > 0$ one has to introduce a jump of the 
trajectory at $\tau = \tau_0$:
\be \label{Evacc3}
 \frac{\partial S(t,\tau)}{\partial t} + \frac{\partial S(t,\tau)}{\partial \tau}
            = - \sigma(\tau) c_0 E(t) S(t,\tau) - u(t,\tau) S(t,\tau) +
           \delta(\tau-\tau_0) \int_{\tau_0}^\infty u(t,s) S(t,s) \dd s,
\ee
where $\delta$ is a unit impulse at zero.\footnote{%
The latter modification requires interpreting the equation in a weak sense.} Of course, it makes sense to suppose that 
$u(t,\tau) = 0$, for all $\tau \in [0,\tau_0)$.  We mention that the last term in the equations
 \reff{Evacc2} and \reff{Evacc3}  does not bring any additional complication
in solving the system numerically.

\paragraph{Measuring vaccination effectiveness through total healthy days.}

To evaluate different vaccination strategies, we adopt a population-level outcome measure based on "healthy days" -- time periods when individuals are susceptible, hence neither infected nor dead. This concept is analogous 
to the "healthy days at home" measure used in healthcare evaluation, see \cite{Burke2020} and
 \cite{Lam-2021}, though, given the model \eqref{EE}--\eqref{EIt0} respectively the variant with vaccination, our metric focuses on infection status rather than location-specific states. 

Other measures like cumulative infections, peak prevalence, or cumulative deaths are present in the literature and are used, dependent on the policy goals of a study. Such measures can be also calculated from the solution of our PDE-model.

However, in the numerical part, which will be purely explorative, we choose healthy days as our primary metric to compare vaccination strategies for several reasons. First, healthy days implicitly captures both infection burden and mortality impact, thus combining morbidity and mortality in a single metric. Second, this measure aligns with health economics perspectives where person-time in good health is a primary outcome, similar to quality-adjusted life years (QALYs) but simplified for our single-disease context. Finally, for comparing vaccination strategies, healthy days per dose administered provides a clear cost-effectiveness metric that is interpretable for policy makers.


For a given vaccination strategy $v := (u^*,\tau^*)$, let $S^v(t,\tau)$ denote the resulting profile of susceptible 
individuals over calendar time and immunity age. The total healthy days over the considered time horizon $T$ 
are given by:
\begin{equation}
    \int_0^{T}\int_0^{\infty} S^v(t,\tau)\dd \tau \dd t.
\end{equation}

This value allows comparison of different vaccination strategies when applied to the same 
initial population and time horizon. 
The efficiency of a vaccination strategy $v$ per dose administered can be measured by:
\begin{equation}
       \text{Eff}_T(u^*,\tau^*)=\frac{\int_0^{T}\int_0^{\infty} \left(S^v(t,\tau)-S(t,\tau)\right)
       \dd \tau \dd t}{u^*\int_0^T\!\! \int_{\tau^*}^\infty S^v(t,\tau) \dd \tau \dd t},
\end{equation}
where $S(\cdot)$ is the susceptible population in the no-vaccination scenario.
This ratio compares the additional healthy days gained (numerator) with the total number of vaccine 
doses administered (denominator), providing a measure of vaccination cost-effectiveness in terms of 
healthy time gained per dose.

\subsection{A model with vital dynamics} \label{SSEModel1}

The basic model \eqref{EE}--\eqref{EIt0} is designed for scenarios where demographic changes may be 
neglected, making it suitable for short time horizons. 
In this framework, natural mortality is assumed to be compensated by births, and the model focuses 
on disease dynamics rather than population demographics. However, for longer-term studies, explicit
 consideration of births and deaths becomes necessary, particularly since deaths due to infection are 
 not directly compensated by additional births, and newborns---contrary to the deceased---lack pathogen-specific immunity.

\begin{figure}
    \centering
    \includegraphics[width=0.75\linewidth]{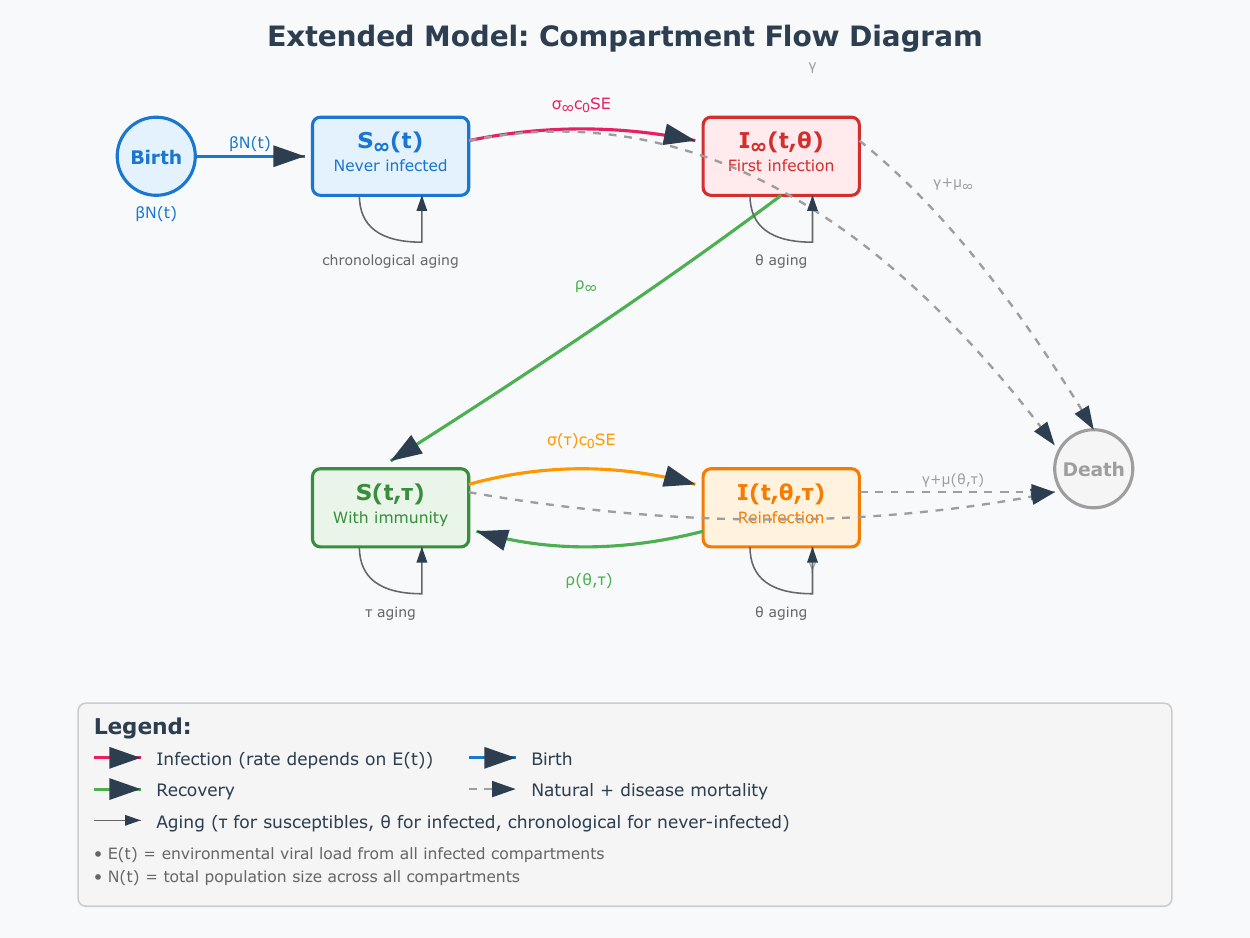}
    \caption{Structure of the extended dual-age model with separated classes for never infected and first-time 
    infected individuals.}
    \label{fig:demograph}
\end{figure}

To this end we may introduce explicit demographic 
processes by adding compartments for individuals who have never been infected: $S_{\infty}(t)$ 
for susceptible individuals with no previous exposure, and $I_{\infty}(t,\theta)$ for individuals 
experiencing their first infection. Additional parameters $\mu_{\infty}(\theta), \rho_{\infty}(\theta),
 \sigma_{\infty}$ must be introduced, representing mortality rate, recovery rate, and susceptibility 
 for first-time infections, respectively. Moreover, note that the introduction of a separate class of fully susceptible individuals allows us to model baseline immunity as different from full susceptibility. The parameter functions $\mu_{\infty}(\theta), \rho_{\infty}(\theta),
 \sigma_{\infty}$ may be treated completely independent of their counterparts $\mu(t,\theta), \rho(t,\theta),
 \sigma$

The extended model creates a four-compartment flow: 
$S_{\infty} \rightarrow I_{\infty} \rightarrow S \rightarrow I \rightarrow S$. 
Individuals who start in $S_{\infty}$, move to compartment $I_{\infty}$ when first infected, 
and then transition to the general compartment of susceptible individuals $S(t,\tau)$ when 
recovering from their first infection. Subsequent infections follow the original dual-age dynamics. 
Figure \ref{fig:demograph} summarizes the described model structure.

This specification allows explicit treatment of births and deaths by adding birth terms and natural 
mortality terms to the respective equations. Here $\beta$ denotes the per capita birth rate, while 
$\gamma$ denotes the natural mortality rate, with births proportional to total population size. 

The extended model reads as follows:

\bda 
\label{ES11}
  \frac{\partial S_{\infty}(t)}{\partial t} 
            &=& \beta N(t) - \sigma_{\infty} c_0 E(t) S_{\infty}(t)  - \gamma S_{\infty}(t),\\
  \frac{\partial S(t,\tau)}{\partial t} + \frac{\partial S(t,\tau)}{\partial \tau}
            &=& - \sigma(\tau) c_0 E(t) S(t,\tau) - \gamma S(t,\tau),\\
  \label{ES011}   S(t,0) &=&  \int_0^{\infty}\!\!\! \int_0^{\infty} \rho(\theta,\tau) I(t,\theta,\tau) 
                  \dd \theta \dd \tau + \int_0^{\infty} \rho_{\infty}(\theta) I_{\infty}(t,\theta) 
                  \dd \theta ,\\
  \frac{\partial I_{\infty}(t,\theta)}{\partial t} + \frac{\partial I_{\infty}(t,\theta)}{\partial \theta}
            &=& - (\gamma + \mu_{\infty}(\theta) + \rho_{\infty}(\theta)) I_{\infty}(t,\theta), \\
  \label{EI11}  \frac{\partial I(t,\theta,\tau)}{\partial t} + \frac{\partial I(t,\theta,\tau)}{\partial \theta}
            &=& - (\gamma + \mu(\theta,\tau) + \rho(\theta,\tau)) I(t,\theta,\tau), \\
             I_{\infty}(t,0) &=& \sigma_{\infty} c_0 E(t) S_{\infty}(t), \\
   \label{EIt011}        I(t,0,\tau) &=& \sigma(\tau) c_0 E(t)S(t,\tau), \\
  \label{EE11}  E(t) &=&  \int_0 ^{\infty}\!\!\! \int_0 ^{\infty} v(\theta,\tau) c(\theta,\tau) I(t,\theta,\tau)
   \dd \theta \dd \tau +  \int_0 ^{\infty} v_\infty(\theta) c_\infty(\theta) I_\infty(t,\theta) \dd \theta,
\eda
and $N(t) = S_{\infty}(t) + \int_0^{\infty} S(t,\tau) \dd \tau + \int_0^{\infty} I_{\infty}(t,\theta) 
\dd \theta + \int_0^{\infty}\int_0^{\infty} I(t,\theta,\tau) \dd \theta \dd \tau$ 
represents the total population size.

When $\beta = \gamma$, natural deaths are compensated by births, recovering the assumption of demographic balance of the basic model. However, all distributions with respect to immunity age will be different from their counterparts in the basic model \eqref{EE}--\eqref{EIt0}, because newborns now start fully susceptible, whereas in the basic model, it is implicitly assumed that the collective of newborn individuals has the same distribution of immunity age as the newly deceased.  

\section{Numerical experiments and comparative analysis} \label{SNumer}

The main purpose of the simulations in this section is to demonstrate the capacity of the model to
produce various types of behavior. This includes dynamics such as convergence to disease free states, endemic states, diminishing or persistent oscillations and more complex behavioral features. This is done for both our basic model and the model with vaccination.

We emphasize that the parameter values or functional forms listed below as a baseline scenario are illustrative rather than empirically estimated. Calibrating the model to specific outbreak data faces fundamental challenges: standard surveillance tracks infection and recovery times but not immunity histories (the value of $\tau$ when individuals become infected); currently, clinical data rarely provide the dual age-dependence required for functions like $\rho(\theta,\tau)$ and $v(\theta,\tau)$; available epidemiological measures (case fatality rates, vaccine efficacy) require substantial statistical transformation to obtain age-dependent functional forms; and the nonlocal term $E(t)$ means only products like $v(\theta,\tau) \cdot c(\theta,\tau)$ are identifiable from transmission data, not individual components. Given these limitations and our primary goal of demonstrating the model's mathematical properties and behavioral repertoire, we choose illustrative, but still biologically motivated, parameters that capture key mechanisms while exhibiting diverse dynamical outcomes.

The unit of time is one day for physical time, infection age and immunity age. We assume

\begin{figure}[!htb]
    \centering
    \begin{minipage}{.50\textwidth}
        \centering
        \includegraphics[width=\linewidth, height=0.20\textheight]{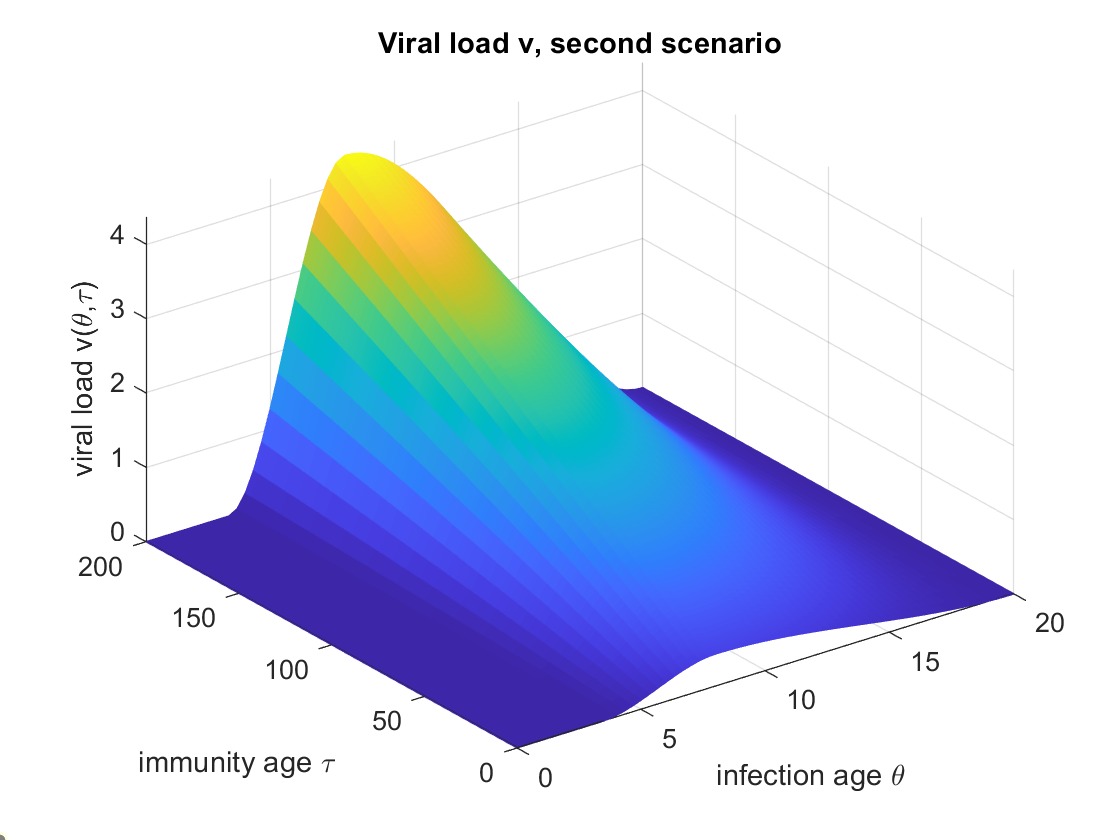}
    \end{minipage}%
    \begin{minipage}{0.50\textwidth}
        \centering
        \includegraphics[width=\linewidth,
    height=0.20\textheight]{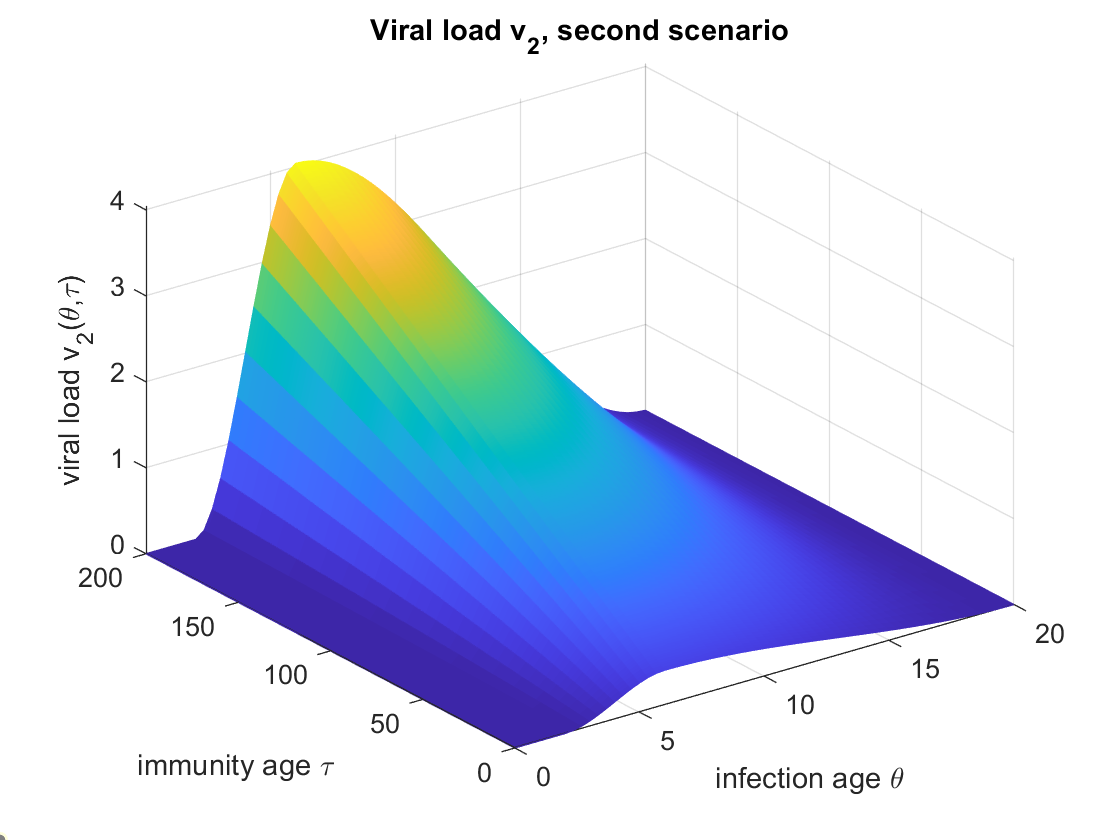}
    \end{minipage}%
    \caption{Two pathogen load functions, $v(\theta,\tau)$ and $v_2(\theta,\tau)$, with the same total load. Both subfigures show the pathogen load depending on infection age $\theta$ and immunity age $\tau$.}
    \label{F_vir}
\end{figure}

\bino
$T = 1200$, \\
$\bar \tau = 200$, \\
$\bar \theta = 20$,\\
$\sigma(\tau) = \min \{0.3, \, 0.3 \tau /\bar \tau \}$, \\
$c_0 = 0.5$, \\ 
$\mu(\theta,\tau) = 0$,\\
$\rho(\theta,\tau) = 0.154$,\\
$c(\theta,\tau) = 2(1- 0.1\,\tau/\bar \tau)(1-0.75\,\theta/\bar \theta)$, \\
$v(\theta,\tau)$ -- see Figure \ref{F_vir}, left plot.
\bino

The above parameters are chosen to be roughly consistent with COVID-19: the maximal immunity period after infection or vaccination is set at 200 days, 
the maximal duration of infection is 20 days, the recovery rate $\rho$
(assumed to be constant for simplicity)\footnote{\label{Fn_rho}%
In fact, the dependence of the recovery rate on $\theta$ and $\tau$ can be rather complicated in reality.} is such that 4.6\% of the infected individuals are still infected on day 20 after infection and recover on this day. 
The function $c(\theta,\tau)$ is specified in a simple way so that it is decreasing in both variables.
The function $c_0 \sigma$ (observe that $c_0$ and $\sigma$
appear in the model only as a product) increases linearly starting from 
zero value (as it makes sense) with a slope $0.15/\bar \tau$. 
In our calculations, we consider no mortality due to the disease in order to avoid shrinking
of the population in the long run, which would overlay the observed effects.
The pathogen load $v$ is obtained by simulation based on the paper  
\cite{LS+PM+NS+VV-24} with clinical data from \cite{Ke+-22}, the later providing  
information concerning $v(\cdot,0)$. The output of the simulation
is in the form of a large table, therefore only the graph is given
on the left plot of Figure \ref{F_vir}.

The functions $c$, $v$, $\sigma$ are determined up to multiplicative constants.
Having in mind the model, the number of constants, including $c_0$ reduces to two,
which in our simulations are chosen 
so that the results are plausible.  In practice, these constants, as well as additional 
shape-parameters for $c$, $v$ and $\sigma$ should be determined using data for the progress of a real disease.

Initial distributions for both susceptible and infected populations across their respective age variables also have to be specified. While these choices are again arbitrary in our non-empirical study, we select initial conditions that represent a biologically plausible scenario of a population with prior disease exposure entering a new epidemic wave.

In the baseline case we assume for $\tau \in [0,\bar \tau]$ 
and $\theta \in [0,\bar \theta]$:
\bda
       s^0(\tau) &=& \tau, \qquad S^0(\tau) = 
   (1-\e)\frac{s^0(\tau)}{\int_0^{\bar \tau} s^0(\tau) \dd \tau},  \qquad \e = 0.01, \\
    i^0(\theta,\tau) &=& 
    \max\left\{0, \,
                  \frac{\tau}{\bar \tau} - \frac{\theta}{\bar \theta}\right\},
       \qquad I^0(\theta,\tau) = 
       \e\frac{i^0(\theta, \tau)}{\int_0^{\bar \theta}\int_0^{\bar \tau} i^0(\theta,\tau) \dd \theta \dd \tau}.
\eda
The total population is normalized to one and the parameter $\e$ represents the overall size of the infected population. In our numerical base case, we set it to $\e = 0.01$.
The functions $s^0$ and $i^0$ define the shapes of the initial data, which are then normalized to $S^0$ and $I^0$.

\bino
In order to avoid showing 3-dimensional figures, we plot aggregated values in most cases, which in particular means that $S$ is shown integrated with respect to $\tau$ for $S$ and $I$ is shown integrated with respect to $(\theta,\tau)$.

\bino
{\bf Numerical experiments with the basic reproduction number(s). 
}

\bino

Using the notations from Section \ref{SBRN}, we have, in general that 
\bd
        \hat \cR_0 \in [0, \cR_0],  
\ed
and by continuity arguments, $ \cR_0[I^0(\cdot)]$ covers all the above interval
when $I^0$ varies over the normalized densities. To demonstrate this numerically, 
we consider a family of distributions parametrized by $\theta' \in [0,\bar \theta]$:
\bd
         I_{\theta^\prime}^0(\tau,\theta) := \delta_{(\theta',\bar \tau)} (\theta,\tau) ,
\ed 
where $\delta_{(\theta',\bar \tau)}$ is the delta-function concentrated at $(\theta',\bar \tau)$.
In the computations we use an approximation with density supported in a domain of size 
0.1$\mbox{day}^2$ in $[0,20]\times [0,200]$. Clearly, we have 
$\cR[I_{\theta'}^0] =  \cR_0$ for $\theta' = 0$.

\begin{figure}[!htb]   
    \centering
    \begin{minipage}{.50\textwidth}
        \centering
    \includegraphics[width=\linewidth, height=0.2\textheight]{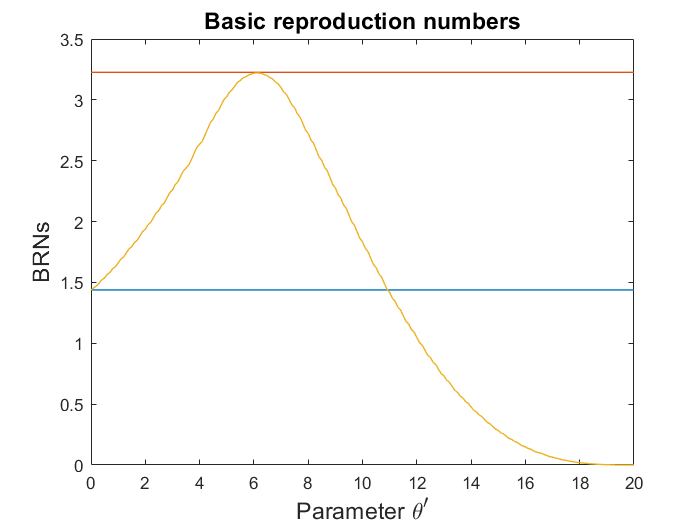}
    \end{minipage}%
    \caption{
Basic reproduction numbers as functions of the parameter $\theta' \in [0,\bar \theta]$: 
(i) $\hat \cR_0$ (corresponding to $\theta' = 0$ and 
$\tau = \bar \tau$ for all the initially infected population) -- lower horizontal line; 
(ii) $\cR_0$ (corresponding to the worst case initial population) -- upper horizontal line;
(iii) $\cR_0[I^0_{\theta'}]$ (corresponding to initial population $I^0_{\theta'}$, $\theta' \in [0, \bar \theta]$) 
-- the curve.
 \label{F_BRN}
}
\end{figure}

As seen on Figure \ref{F_BRN}, the worst case BRN $\cR_0$ is (approximately) attained for $I^0_{6.1}$ and, as expected, all values between zero and $\cR_0 \approx 3.226$ are attained by $\cR_0[I^0]$ for
some initial distribution $I^0$. It is remarkable that the range of $\cR_0[I^0]$ is large, also
including values smaller or bigger than one. The realization of the worst case, however, 
is hard to happen in practice. It seems reasonable to use $\hat \cR_0$ as a more realistic upper
bound for the BRN, in our case $\hat \cR_0 \approx 1.439$. 

\bino
{\bf Numerical experiments with the basic model.} 

\begin{figure}[!htb]
    \centering
    \begin{minipage}{.50\textwidth}
        \centering
        \includegraphics[width=\linewidth, height=0.2\textheight]{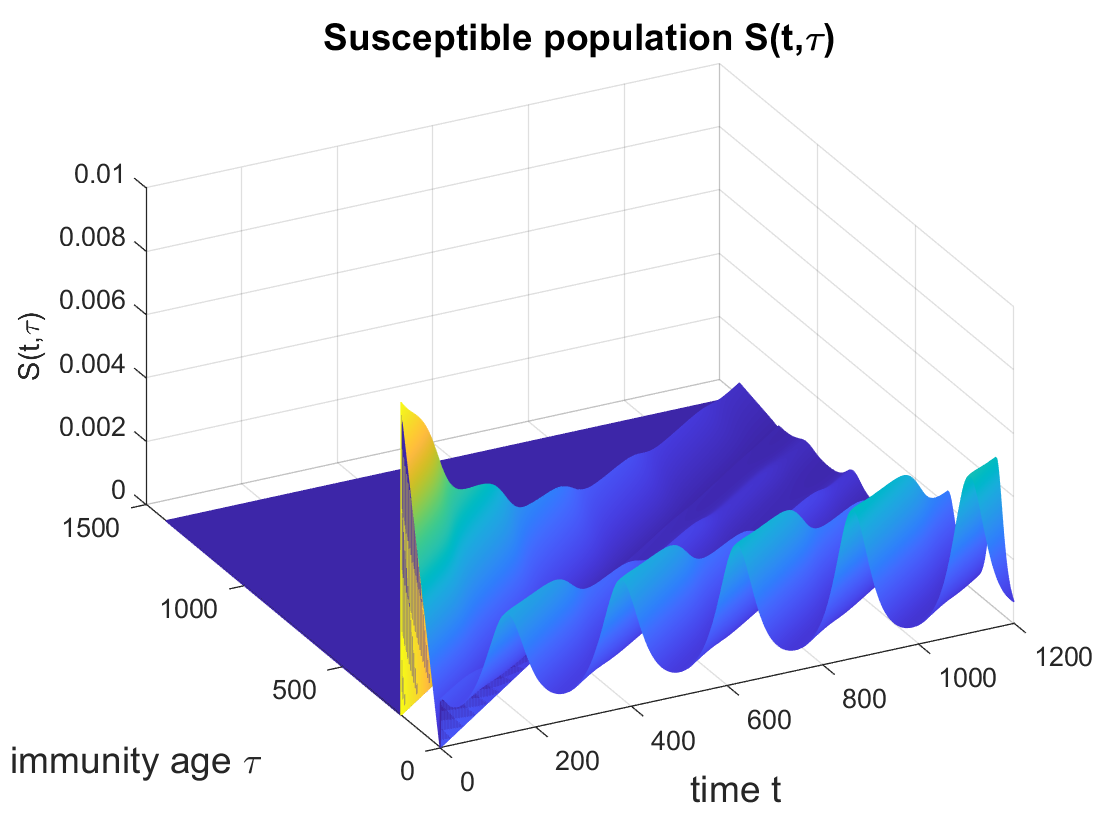}
    \end{minipage}%
\begin{minipage}{0.50\textwidth}
        \centering
        \includegraphics[width=\linewidth, height=0.2\textheight]{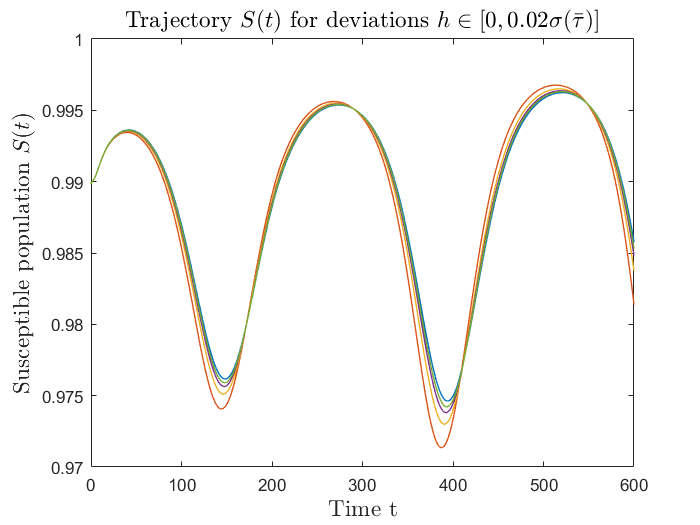}
    \end{minipage}%
    \caption{Left plot: the susceptible population  $S(t,\tau)$ for the baseline case.
The figure shows the time $t$ dependent evolution of the distribution of 
the susceptible population $S(t, \tau)$ w.r.t. the immunity age $\tau$.
Right plot: the susceptible population for various scenarios for the deviation
     $\Delta(\tau) := \sigma(\tau) - \sigma(\bar \tau)$ for $\tau \geq \bar \tau$;
      the lines correspond to $h = \|\Delta \|_{L_\infty}$ as percentage of $\sigma(\bar \tau): 
   \;2, \,1, \,0.5, \,0.25, \,0$.}
    \label{F_3DS}
\end{figure}

\bino
{\em Observation 1.}
A 3-dimensional plot of $S(t,\tau)$ in the baseline case for model \eqref{EE}--\eqref{EIt0} is shown in Figure \ref{F_3DS}. It demonstrates the dynamics of the immunity age distribution over time. In particular, one can see the waves induced by the combined effect of immunity waning and reinfections.

\bino
In the next few paragraphs, we investigate how cutting the immunity age horizon to $[0,\bar \tau]$ influences
the evolution of the modeled disease.  So far we assume that the $\tau$-dependent parameters
keep their value at $\bar \tau$ for $\tau > \bar \tau$. We focus on the most critical parameter,
$\sigma$. Keeping the notation $\sigma(\tau) = \sigma(\bar \tau)$ for $\tau > \bar \tau$, 
we assume that the ``true'' value of $\sigma$ is an unknown $\tilde \sigma(\tau)$ for 
$\tau > \bar \tau$. Denote by $(S[\tilde \sigma](t), I[\tilde \sigma](t))$ the  corresponding 
solution of \eqref{EE}--\eqref{EIt0}, aggregated in $\tau$, resp. $(\theta,\tau)$. 
A reasonable measure of sensitivity of the solution with respect to $\sigma$, 
hence with respect to $\bar \tau$, is a pair of numbers $(L_S,L_I)$ (if it exists) such that 
\bd
              \frac{1}{T} \int_0^T | S[\tilde \sigma](t) - S[\sigma](t) |   \dd t 
         \leq L_S \| \tilde \sigma - \sigma \|_\infty     
\ed
for every measurable and bounded $\tilde \sigma$ with $\| \tilde \sigma - \sigma \|_\infty$ --
sufficiently small. Here and below $\| \cdot \|_\infty$ and $\| \cdot \|_1$ are the norms
in the spaces $L_\infty$ and $L_1$, respectively, on an appropriate scalar interval.
Similarly for $I$. Equivalently,
\bd
     L_S = \lim_{h \ra +0} \frac{1}{hT} \sup_{\| \psi \|_\infty = 1}   
        \int_0^T | S[\sigma + h \psi](t) - S[\sigma](t) |   \dd t. 
\ed
After choosing a sufficiently representative set of unit functions $\psi(\tau)$ we compute for small $h > 0$ the value of the expression
after ``$\lim$" in the last exposed formula (call it $L_S(h)$).
Similarly we define and compute $L_I(h)$.
On Figure \ref{F_3DS} one can see the trajectories $S$ corresponding to
$h = 0.006, 0.003, 0.0015, 0.00075$. The values of $L_S(h)$ and $L_I(h)$ 
for these values of $h$ (and also for $h = 0.000375$) are given on Table \ref{TaL}. 
The important message is that the numbers $L_S(h)$ and $L_I(h)$ behave as 
uniformly bounded as $h \ra 0$, which gives evidence that the calmness constants 
$L_S$ and $L_I$ exist and one can estimate their values.  

\begin{table}[h]
\begin{center}
\begin{tabular}{c | c  c c  c c }
                     $h$    & 0.006 & 0.003 & 0.00015 & 0.00075 & 0.000375\\ 
 \hline
 $L_S(h)$ & 0.2005 & 0.1984 & 0.1973 & 0.1967 & 0.1964 \\
 $L_I(h)$ & 0.2042 & 0.2021 & 0.2010 & 0.2004 & 0.2001 
\end{tabular}
\caption{Shows boundedness of the values of the $L_S(h)$ and $L_I(h)$ 
for various small values of $h$. }
\label{TaL}
\end{center}
\end{table}

\begin{figure}[!htb]
    \centering
    \begin{minipage}{.50\textwidth}
        \centering
        \includegraphics[width=\linewidth, height=0.2\textheight]{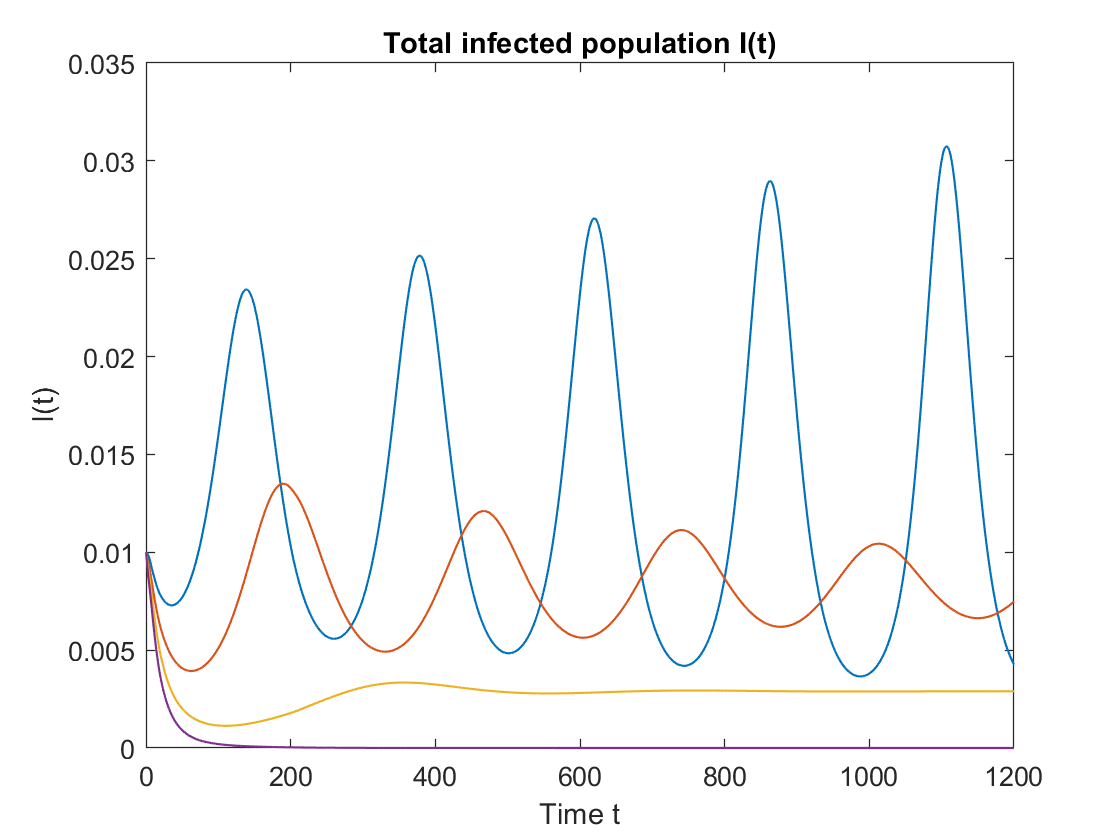}
    \end{minipage}%
    \begin{minipage}{0.50\textwidth}
        \centering
        \includegraphics[width=\linewidth, height=0.2\textheight]{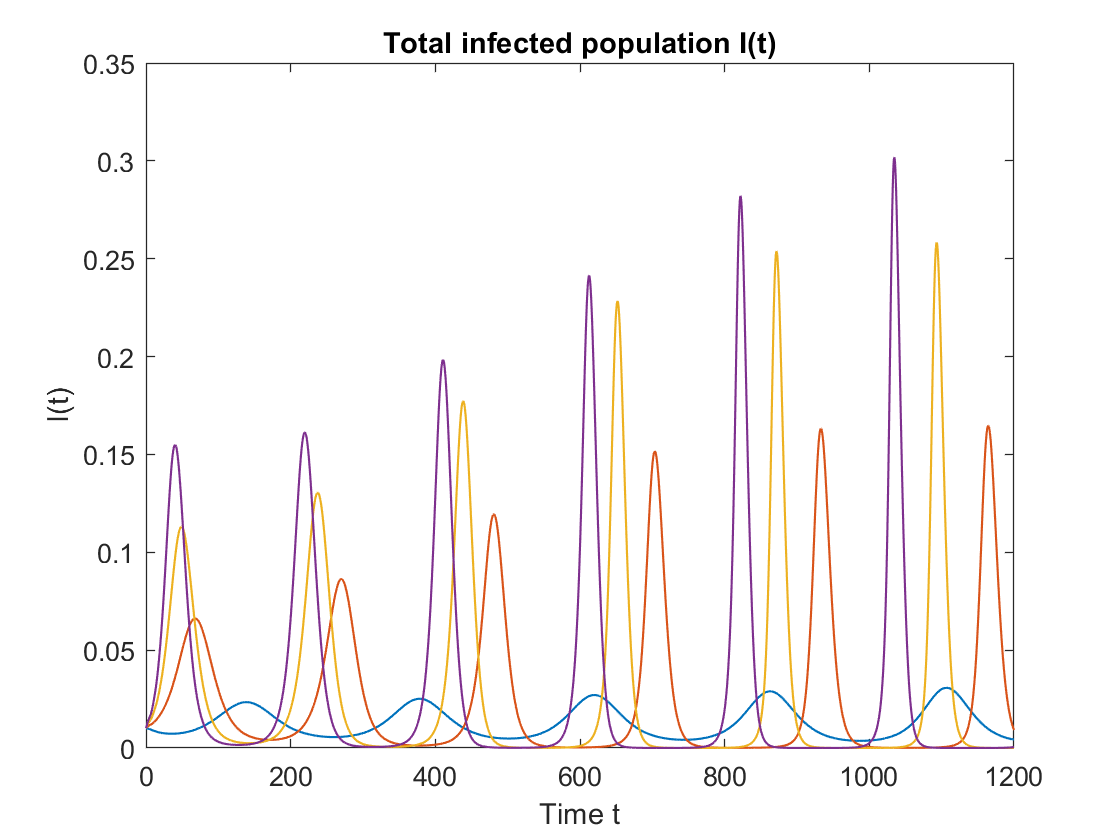}
    \end{minipage}%
    \caption{Total infected population for various values of the parameter $c_0$ in the baseline setting:
    $c_0 = 0.50 \text{ (blue)}, \, 0.45 \text{ (red)}, \, 0.40 \text{ (yellow)}, \, 0.35 \text{ (magenta)}$ on the left plot, and $c_0 = 0.5 \text{ (blue)}, \, 0.7 \text{ (red)}, \, 0.9 \text{ (yellow)}, \, 1.1 \text{ (magenta)}$ on
    the right plot. Note that the reference $c_0 = 0.50$ is blue in both sub-figures. In all cases curves with larger oscillations correspond to higher values of $c_0$.}
    \label{F_Ic0}
\end{figure}

For comparison, Figure \ref{F_Ic0} represents the evolution of the infected populations for several values  
of the contact rate $c_0$, namely for $c_0 = 0.50, \, 0.45, \, 0.40, \, 0.35$ on the left plot, and 
$c_0 = 0.5, \, 0.7, \, 0.9, \, 1.1$ on the right plot. As expected, higher values of $c_0$ correspond to curves with larger oscillations. We see on the left plot that 
the magnitude of the oscillation increases with time when $c_0 = 0.5$, while it decreases for $c_0 \leq 0.45$ and even $I(\cdot)$ becomes monotonically decreasing to zero for $c_0 = 0.35$. It makes sense to expect that for some value of $c_0 \in (0.45, 0.5)$ the trajectory is periodic. However, a formal proof consistent with this observation remains open. 

The increase of the magnitude of the oscillations is also visible on the right plot. It is remarkable that in both plots higher values of $c_0$ lead to more
steepness of the curves, combined with longer periods where $I(t)$ is close to zero. To explain 
this effect, assume that $c_0$ is ``very large''. Then after a short time, ``very many'' people will be 
infected and as a result, after a some time $t < \bar \tau= 200$ ``very many'' people will have high
immunity. Since $\sigma(0) = 0$, thus $\sigma(\tau)$ is small for small $\tau$, 
the flux of new infections will be ``very small'' (actually the effective reproduction number 
will be much smaller than one) and the infected
population will quickly become ``very small''\footnote{\label{Fn_BRN}%
Using the representation \reff{ERN_SI}, one can see that the BRN is small when $S^0$ is 
mainly concentrated close to zero. This is due to $\sigma(0) = 0$.}. 
After awhile, the immunity age will increase enough for a new cycle to
appear. The next observation provides support of this mechanism.

\begin{figure}[!htb]
    \centering
    \begin{minipage}{.50\textwidth}
        \centering
        \includegraphics[width=\linewidth, height=0.2\textheight]{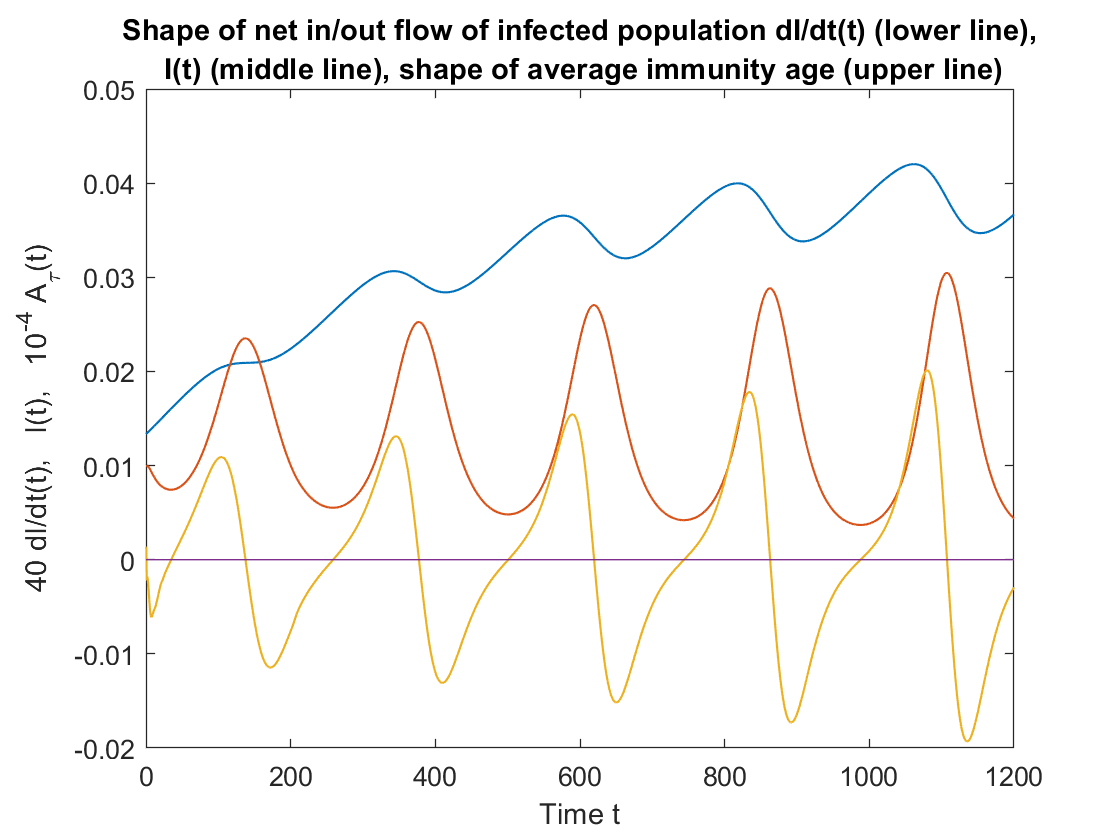}
    \end{minipage}%
    \caption{Instantaneous change of the size of the infected population, $\frac{\dd I}{\dd t}(t)$
    scaled by 40 (lower curve, yellow),
    infected population $I(t)$ (middle curve, red), and the average immunity age 
     $A_\tau(t)$ scaled by $10^{-4}$ to fit on the same plot (upper curve, blue). }
    \label{F_Av}
\end{figure}

\bino 
{\em Observation 2.}
Figure \ref{F_Av} shows three curves computed for the baseline specification. 
The middle one is the trajectory of the infected population $I(t)$. 
The lower line is the instantaneous change of $I$, that is, $\frac{\ddd I}{\ddd t}(t)$
multiplied by 40 (for visual comparison of the three curves).
The upper line is the average immunity age of the susceptible population, $A_\tau(t)$, 
multiplied by $10^{-4}$ (again for comparison). 
The average immunity age is defined as usual:
\bd
           A_\tau(t) := \frac{\int_0^{T+\bar \tau} \tau S(t,\tau) \dd \tau}{S(t)}.
\ed
Only the shapes of the curves matter. 

One can clearly see that times $t$ with lower $A_\tau(t)$ (meaning high immunity)
lead to negative $\frac{\ddd I}{\ddd t}(t)$ and the infected population shrinks. This observation
is consistent with formula \reff{ERN_SI} for the basic reproduction number. Indeed,  
Footnote \ref{Fn_BRN} means that low average immunity age leads to low BRN, which is equivalent
to local increase of the susceptible population (in the case of zero mortality considered in the 
baseline case) and to decrease of the infected population. 

Observation 2 confirms the mechanism described in Observation 1: periods of low average immunity age (high population immunity) directly correspond to negative dI/dt, causing the infected population to shrink.

\begin{figure}[!htb]
    \centering
    \begin{minipage}{.50\textwidth}
        \centering
        \includegraphics[width=\linewidth, height=0.2\textheight]{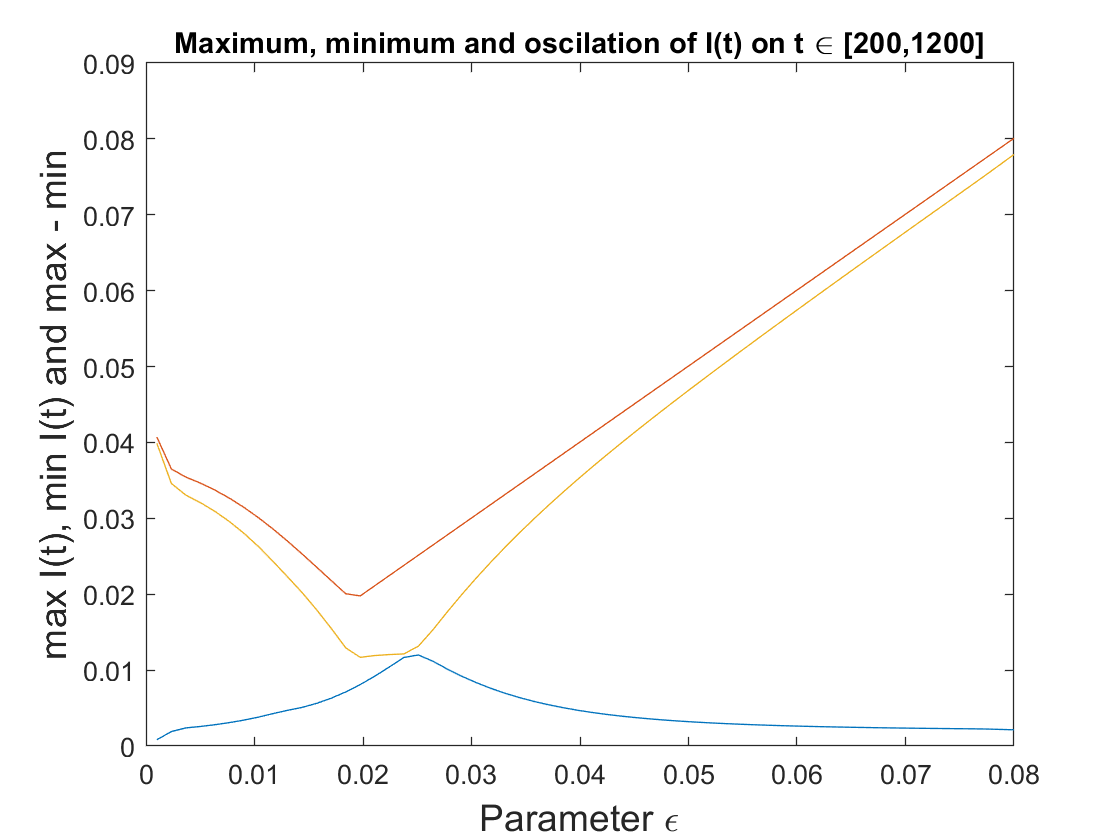}
    \end{minipage}%
    \begin{minipage}{0.50\textwidth}
        \centering
        \includegraphics[width=\linewidth, height=0.2\textheight]{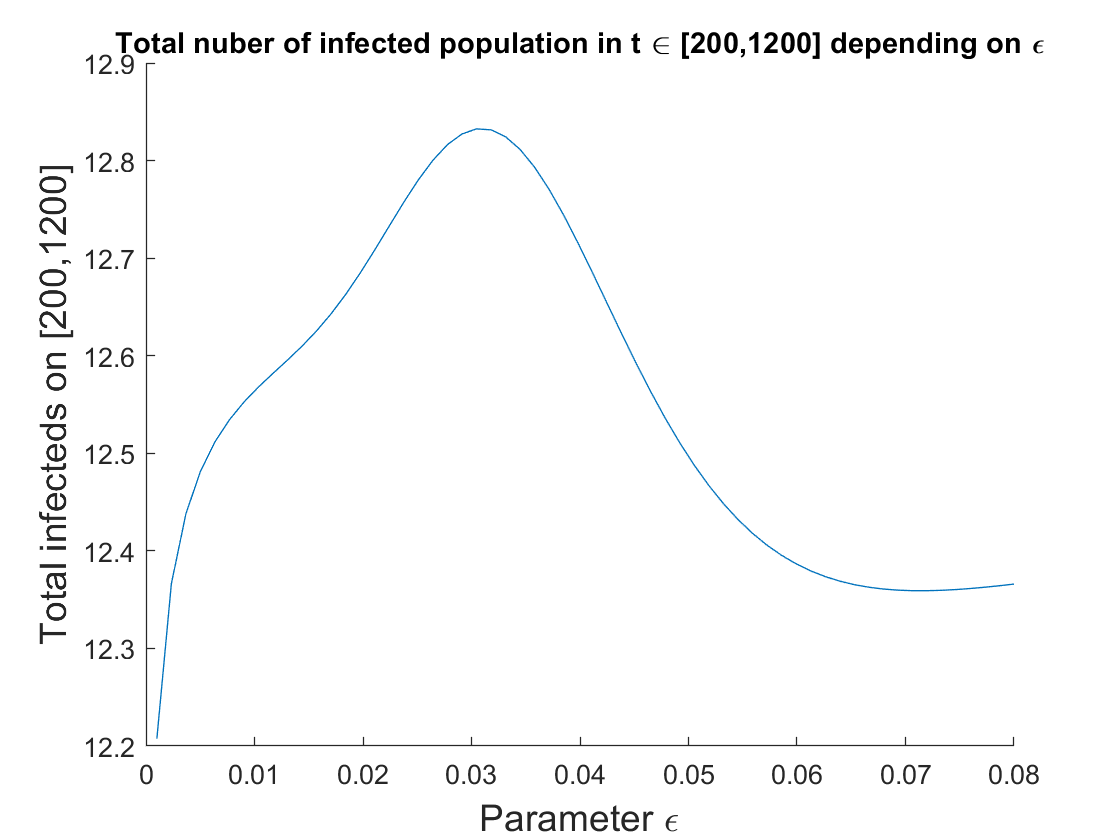}
    \end{minipage}%
    \caption{For the baseline case, the size of the initially infected population, $\e$, varies from
    $0.001$ to $0.08$.  Left plot: $\min_{t\in [\bar \tau,T]} I(t)$ (lower line, blue),  
    $\max_{t\in [\bar \tau,T]} I(t)$ (upper line, red) and the oscillation of $I$ (intermediate line, yellow) for various 
    $\e$.  Right plot: $\int_{\bar \tau}^T I(t) \dd t$ for various $\e$.}
    \label{F_Osc}
\end{figure}

\bino 
{\em Observation 3.}
The next experiment investigates the dependence of the trajectories of $S$ and $I$ on the initial data.
Only the size of the initially infected population varies from $\e = 0.001$ to $\e = 0.08$,
while the profile of the $I^0$ and $S^0$ stays the same. For each $\e$, the maximal value, 
the minimal value, and the oscillation of the corresponding trajectory $I(\cdot)$ on $[\bar \tau,T]$ 
are computed. The corresponding lines are presented on the left plot of Figure \ref{F_Osc}.
The initial interval $[0, \bar \tau]$ is ignored in order to capture the long run behavior. 
The intermediate curve represents the oscillation
\bd
   \max_{t\in [\bar \tau,T]} I(t) - \min_{t\in [\bar \tau,T]} I(t). 
\ed
It is interesting that each of the plotted function is not monotonic. For example, 
minimal oscillation is attained for $\e$ between $0.018$ and $0.025$. 
The right plot represents the total number of infected individuals, $\int_{\bar \tau}^T I(t) \dd t$,
for different initial sizes $\e$. The behavior is also non-monotonic. The main conclusion
here is that there is no direct relation between the size of the initially
infected population, on one hand, and the oscillation and the total number of infected
individuals, on the other hand.

\bino
{\bf Experiments involving vaccination.} 

\bino
First, we compare the results for two different pathogen load functions, $v$ and $v_2$,
(see Figure \ref{F_vir}). The second one differs from the baseline scenario $v$ in that the pathogen load 
is higher at low infection ages and lower at higher infection ages, while the total pathogen load is the same.

\begin{figure}[!htb]
    \centering
    \begin{minipage}{.50\textwidth}
        \centering
        \includegraphics[width=\linewidth, height=0.2\textheight]{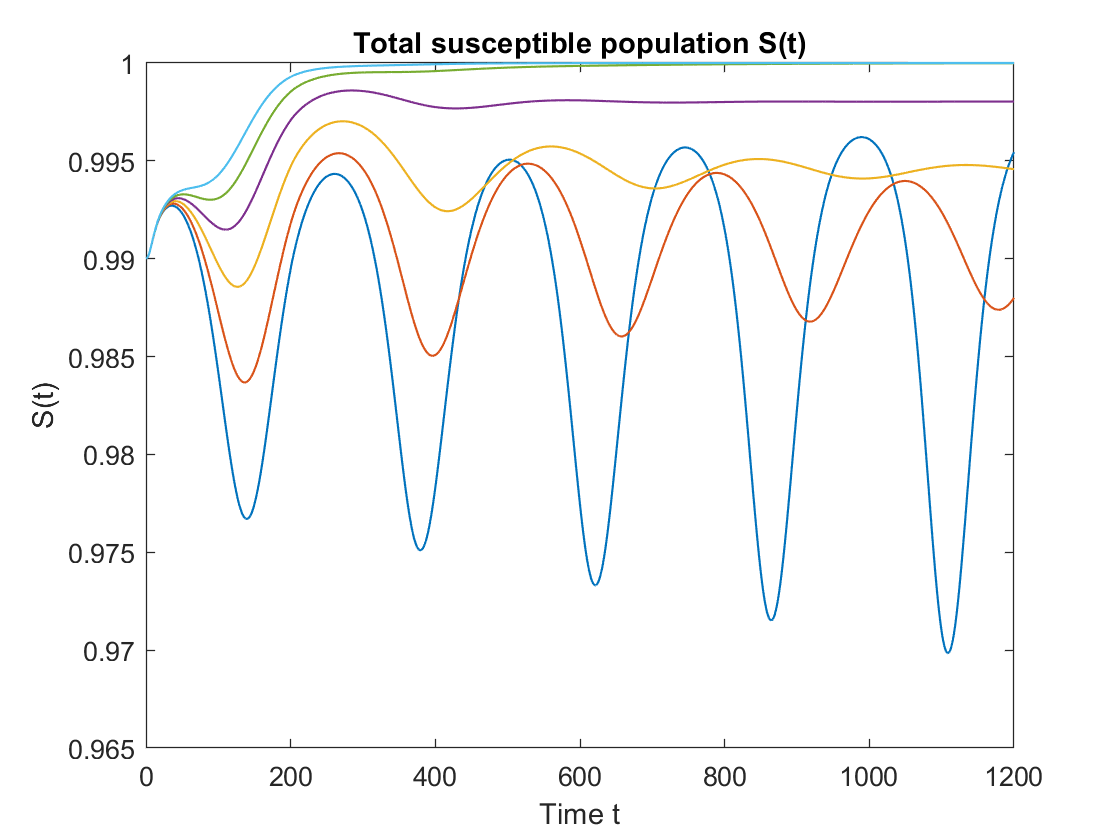}
    \end{minipage}%
    \begin{minipage}{0.50\textwidth}
        \centering
        \includegraphics[width=\linewidth, height=0.2\textheight]{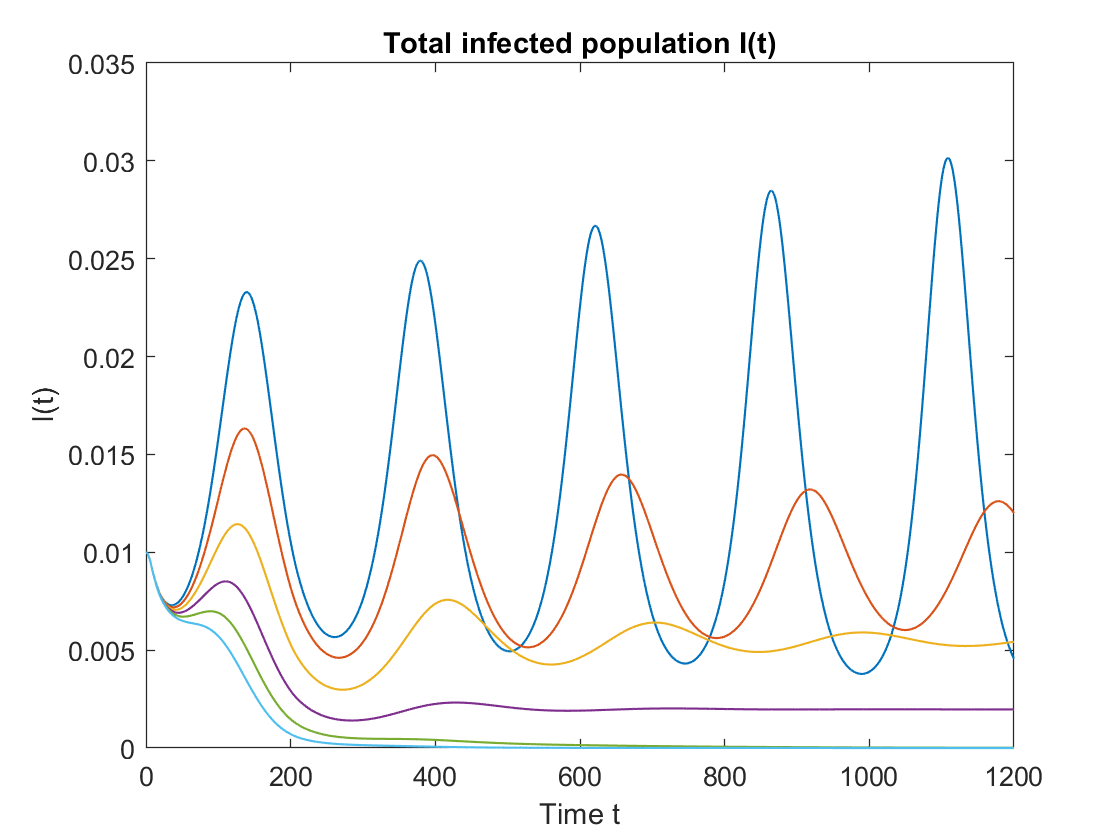}
    \end{minipage}%
    \caption{For the baseline pathogen load $v$: susceptible (left) and infected (right) populations 
    without and with vaccination. The case without vaccination is shown in dark blue. Additionally, we show the development for the five vaccination rates $u^*$ : 0.001 (red), 0.002 (yellow), 0.003 (magenta), 0.004 (green), 0.005 (light blue), which roughly corresponds 
    to vaccinating between 10\% and 50\% of the susceptible population with $\tau > \bar \tau$ in 100 days.}
    \label{F_SI}
\end{figure}

\begin{figure}[!htb]
    \centering
    \begin{minipage}{.50\textwidth}
        \centering
        \includegraphics[width=\linewidth, height=0.2\textheight]{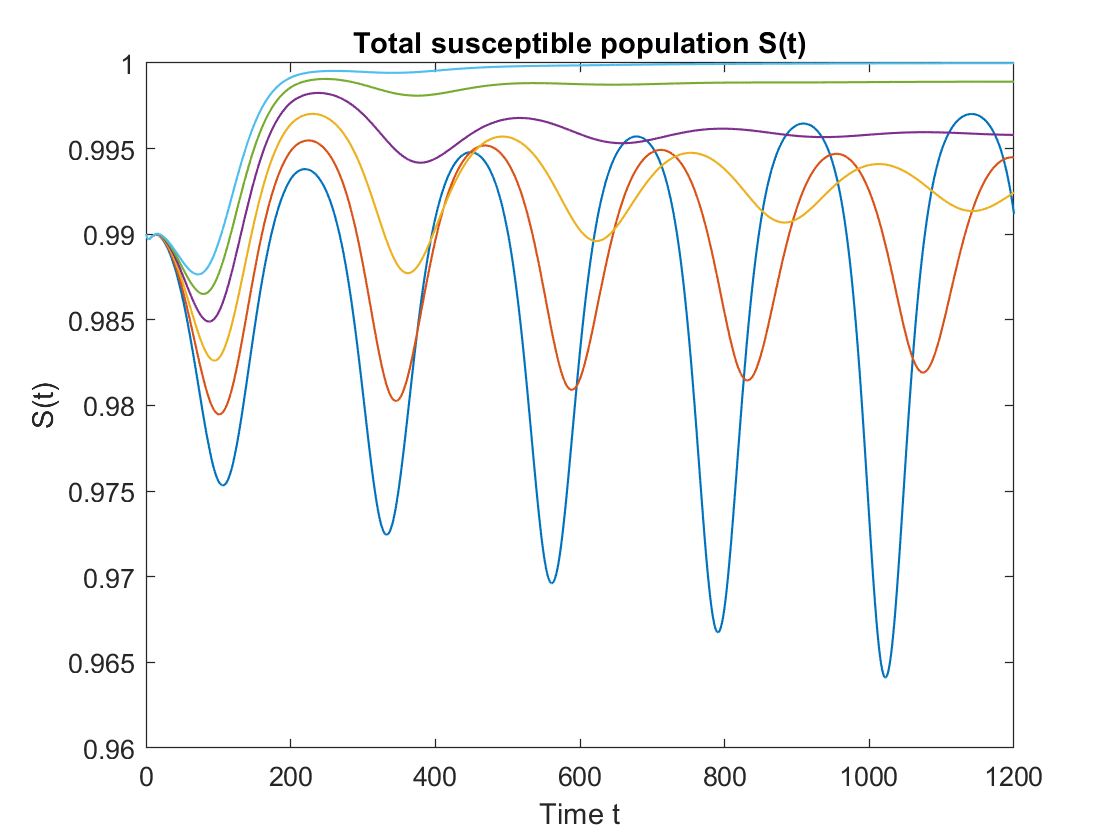}
    \end{minipage}%
    \begin{minipage}{0.50\textwidth}
        \centering
        \includegraphics[width=\linewidth, height=0.2\textheight]{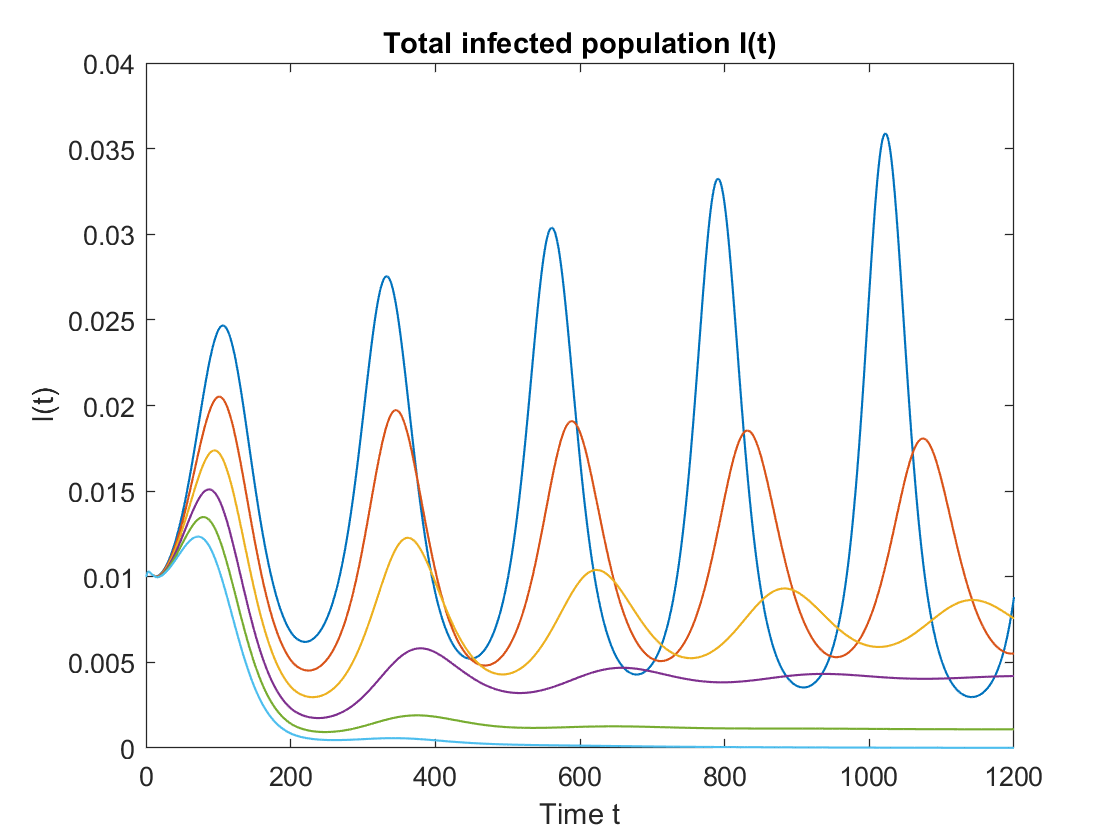}
    \end{minipage}%
    \caption{For the alternative pathogen load $v_2$: susceptible (left) and infected (right) populations 
    without vaccination and with vaccination. The case without vaccination is shown in dark blue. Additionally, we show the development for the five vaccination rates $u^*$ : 0.001 (red), 0.002 (yellow), 0.003 (magenta), 0.004 (green), 0.005 (light blue), which roughly corresponds 
    to vaccinating between 10\% and 50\% of susceptibles with $\tau > \bar \tau$ in 100 days.}
    \label{F_SI_2}
\end{figure}

\bino
{\em Observation 4.} Figure~\ref{F_SI} and Figure~\ref{F_SI_2} represent the total size of susceptible and infected
subpopulations resulting from six vaccination rates $u^* \in \{0, \, 0.001, \, 0.002, \, 0.003, \, 0.004,
\, 0.005\}$ applied to people with immunity ages above $\tau^* = \bar \tau = 200$. 
In both scenarios for $v$, higher lines for $S(t)$ and lower lines for $I(t)$ correspond to higher
vaccination rate $u^*$. As seen in figures \ref{F_SI} and \ref{F_SI_2}, the pathogen load function $v_2$, 
where the distribution is shifted to lower infection ages, appears as more aggressive. 
In particular, in the case of pathogen load function $v$, vaccination rate about $u^* \approx 0.003$ leads to
endemic state and for higher vaccination rates the disease dies out. In the case of pathogen 
load function $v_2$, the same happens for a higher value $u^* \approx 0.004$. Similar observations
were made in all our experiments of this kind. 

\begin{figure}[!htb]
    \centering
    \begin{minipage}{.50\textwidth}
        \centering
        \includegraphics[width=\linewidth, height=0.18\textheight]{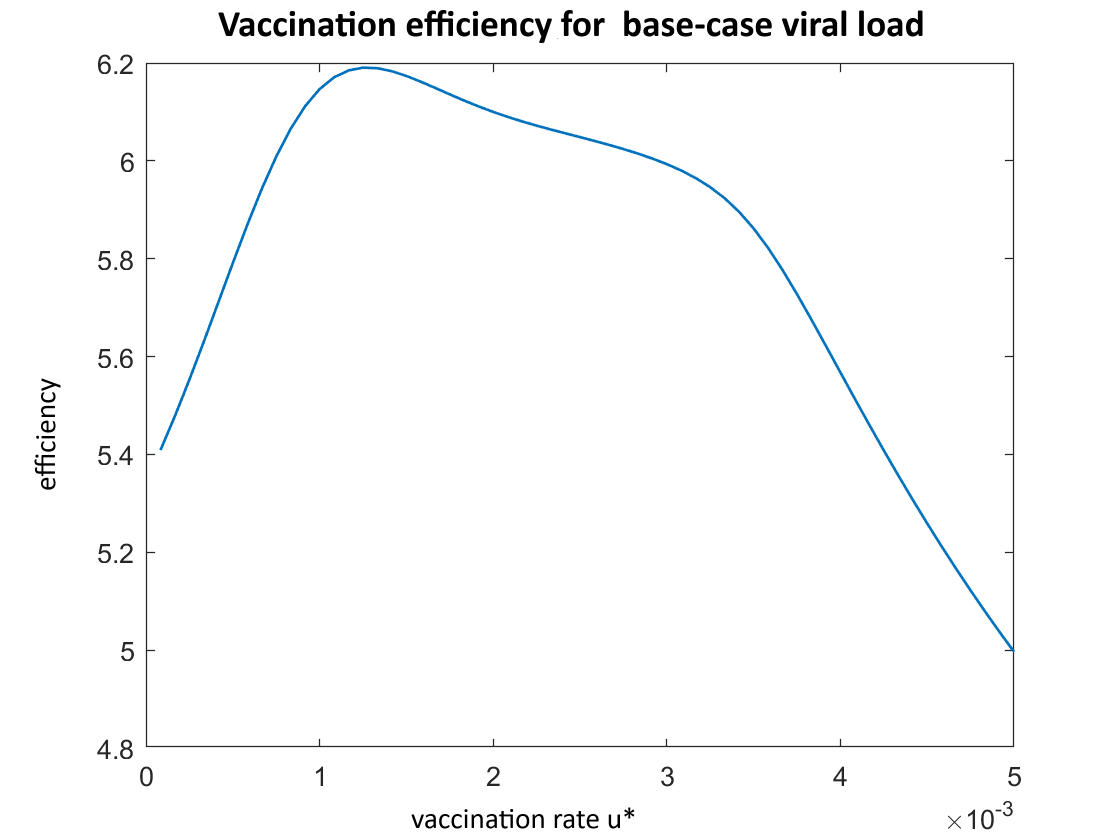}
    \end{minipage}%
    \begin{minipage}{0.50\textwidth}
        \centering
        \includegraphics[width=\linewidth, height=0.18\textheight]{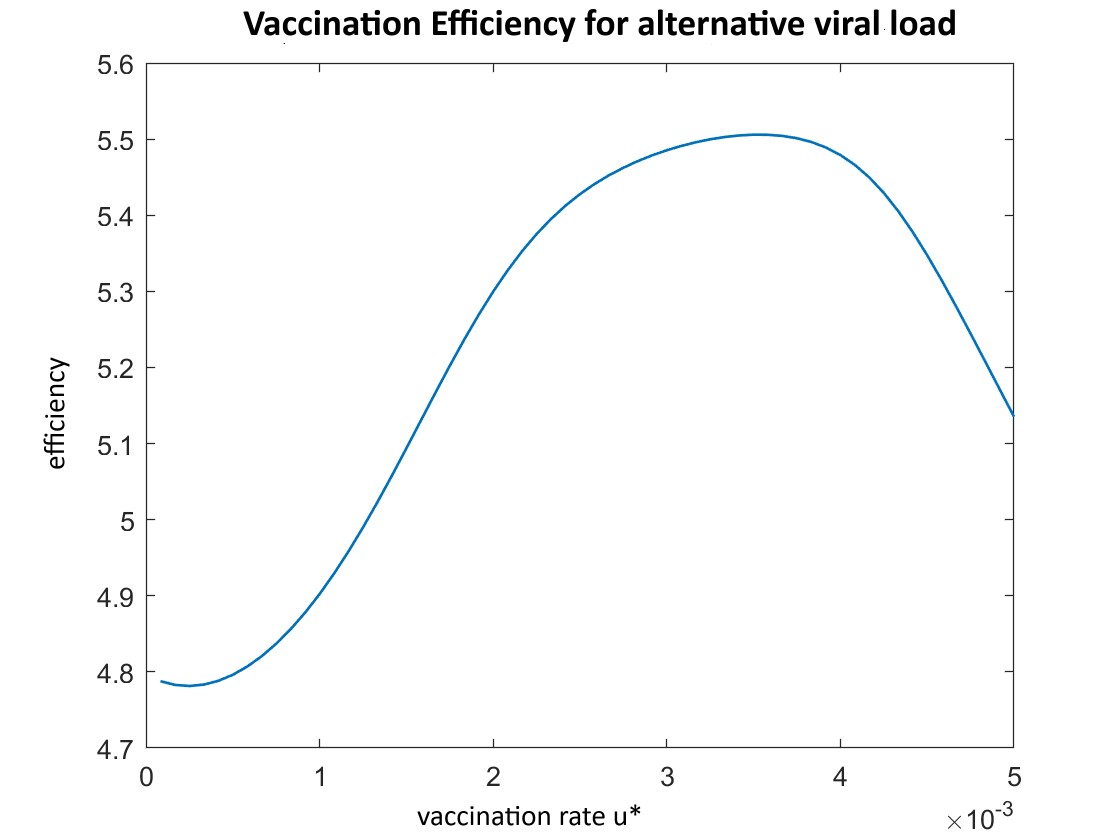}
    \end{minipage}%
    \caption{Efficiency of vaccination for $u^* \in [0.0002,0.005]$, corresponding to the
    pathogen load functions $v$ (left plot) and $v_2$ (right plot).}
    \label{F_Eff}
\end{figure}

\bino
{\em Observation 5.} The left plot of Figure \ref{F_Eff} represents the efficiency of vaccination when 
applied to immunity ages 
above $\tau^* = \bar \tau = 200$, with vaccination rates varying between 0.0002 and 0.005.  
The two plots correspond to pathogen load functions $v$ (left plot) and $v_2$ (right plot). 
The most efficient vaccination rate shifts to higher value for the more aggressive pathogen load profile.
The situation is similar if the contact rate or the susceptibility $\sigma$ increase: more danger leads to
higher most efficient vaccination rate. 

\begin{figure}[!htb]
    \centering
    \begin{minipage}{.50\textwidth}
        \centering
        \includegraphics[width=\linewidth, height=0.2\textheight]{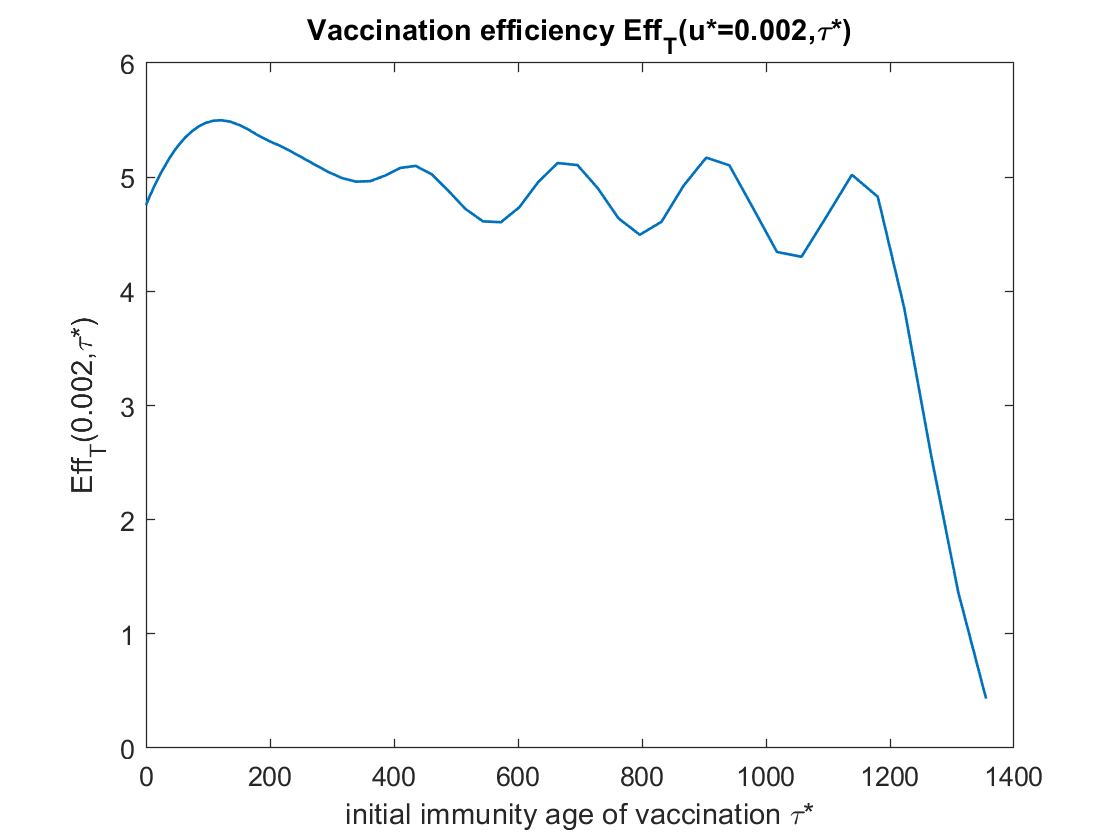}
    \end{minipage}%
    \begin{minipage}{0.50\textwidth}
        \centering
        \includegraphics[width=\linewidth, height=0.2\textheight]{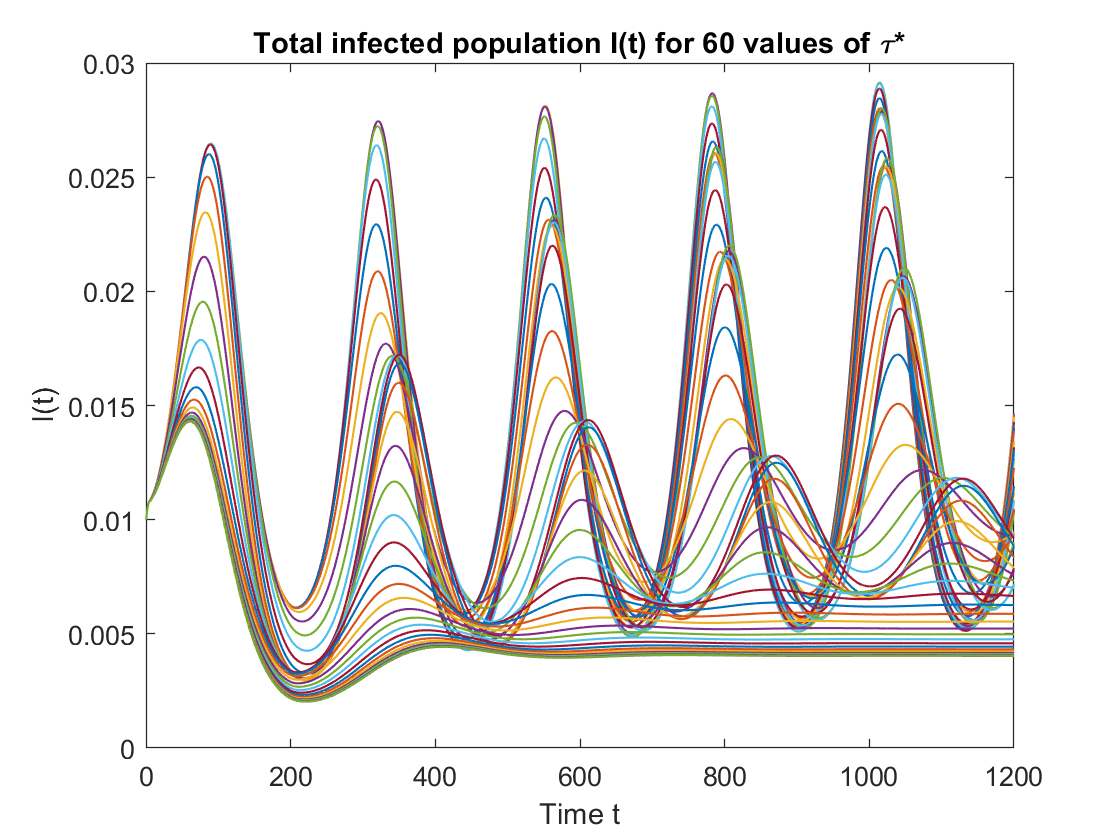}
    \end{minipage}%
    \caption{The left plot represents the efficiency of vaccination with rate $u^* = 0.002$ for vaccination at 60 possible immunity ages $\tau^*$ between $0$ and $T+\bar\tau = 1400$, corresponding to the pathogen 
    load function $v_2$. The right plot shows
    the corresponding trajectories of the infected subpopulation.}
    \label{F_Eff_2}
\end{figure}

\bino
{\em Observation 6.}  The left plot of Figure \ref{F_Eff_2} represents the efficiency of 
vaccination with rate $u^* = 0.002$ for 60 different 
values of $\tau^* \in [0,1400]$, corresponding to the pathogen load function $v_2$. The right plot gives
the 60 trajectories of the infected subpopulation.
Interestingly, the efficiency of vaccination exhibits similar oscillation behavior as the size 
of susceptible and the infected subpopulations, although the oscillations are small. The latter 
excludes vaccination of individuals with immunity age above 1200 days, consistently with  
the time horizon of interest, $T=1200$.    

The most efficient vaccination begins at 
immunity age $\tau^* = 122$. Another interesting 
observation is that with the given intensity of vaccination, $u^* = 0.002$, the epidemic does not
extinct even when applied to all individuals (that is, $\tau^* = 0$). The computation shows that 
the threshold vaccination rate (that is, the minimal vaccination rate for which the disease dies out
when applied with an appropriate minimal immunity age) 
for our data specification is $\hat u^* \approx 0.0028$.  

\section{Discussion} \label{SDiscus}

The aim of this work is to provide a comprehensive description of the dynamics governing an individual’s immune status and their current infection stage simultaneously. To achieve this, we develop a dual-age structured modeling approach that explicitly tracks individuals based on two key metrics: the time elapsed since their last recovery from infection (immunity age) and the time elapsed since their current infection (infection age). We formulate the mathematical structure of the model and derive an expression for the total infection pressure encountered by susceptible individuals. 
 In addition, we calculate the basic reproduction number, which defines the conditions under which an epidemic may emerge. 

A notable extension of the model allows for vaccination based on immunity age. This feature enables the development of targeted vaccination strategies, where individuals can be prioritized based on the time elapsed since their last vaccination or infection. This approach may optimize the allocation of vaccination resources for effective disease control.
Another extension incorporates demographic dynamics such as births and deaths, which influence long-term epidemiological trends.

Our numerical experiments reveal several important insights. Higher contact rates produce larger amplitude oscillations with longer inter-epidemic periods, driven by immunity age dynamics in the population. The relationship between initial infection levels and long-term epidemic behavior is non-monotonic, indicating complex dependence on initial conditions. The average immunity age of the susceptible population strongly influences pathogen transmission dynamics, with lower average immunity age (higher population-level immunity) suppressing infection growth. Vaccination
efficiency depends critically on the initial pathogen load profile, with more pronounced pathogen load distributions (shifted toward earlier infection ages) requiring higher vaccination rates for disease elimination. Most efficient vaccination strategies are often achieved at intermediate immunity ages rather than targeting only individuals with fully waned immunity. 

The work presented here focuses on the mathematical elaboration of a new modeling framework. Estimation of hazard rates with dependence on more than one time variable is not standard practice in clinical statistics and it is difficult to find corresponding data or results in the literature. Empirical calibration of the model is, therefore, left for future research. In principle, estimation is possible from public health microdata, involving points in time of infection (tested) and recovery for individual infection episodes. Infected individuals are tracked by an anonymized key. This type of data---although usually not publicly 
available---were collected in many countries during the Covid-19 pandemic; see, e.g., \cite{StatistikAustria-AMDC}. Together with statistical information about the total population size, those data contain the relevant information to apply standard survival regression methods to estimate the hazard rates $\rho$ and $\mu$, when treating the second time variable $\tau$ as a covariate. Moreover it is possible to estimate the products $v(\theta,\tau)\cdot c(\theta, \tau)$ and $c_0\cdot \sigma(\tau)$, which requires more specialized methods than classical hazard rate estimation: The infection pressure $E(t)$ is given by an integral over the infected population weighted by a kernel function, where we integrate over time since infection and immunity age. Here, the kernel function must be estimated, necessitating functional data analysis, e.g., penalized spline estimation for function-on-function regression \cite{Rao+2023}.

In our numerical calculations, for simplicity, we treated immunity obtained from infection and vaccination 
as being identical. In reality, vaccine-induced immunity may differ in duration or effectiveness compared to infection-acquired immunity. 
We described a natural extension in Section~\ref{SSVacMod} for the case that vaccination provides less immunity than infection, possibly dependent on the current 
$\tau$ at the time of vaccination. This requires introducing either an additional parameter function or 
an impulse term. 
Although mathematically more complex, this extension would allow more realistic modeling of different types of immunity and their relative effectiveness. 
It is also possible to extend the model with different parameter functions for naturally recovered and vaccinated individuals. However, this would afford the introduction of a new heterogeneity variable -- time since last vaccination.

Like other deterministic epidemiological models, our framework might be extended to incorporate additional heterogeneity or stochasticity.

Regarding contact structure heterogeneity, the proposed model (like any other ODE or PDE model) can serve as building block for network or metapopulation models. Nodes could represent geographical regions, age groups, or subpopulations with varying vulnerability, each related to similar system of equations with distinct parameter functions $v(\theta,\tau), \sigma(\tau), \rho(\theta,\tau)$, etc. The coupling between nodes requires additional transmission terms representing between-group contacts. Importantly, immunity-age structure may interact with network topology: For example, individuals at highly connected nodes might experience more frequent reinfections, and also affect the distribution of immunity ages across other network nodes.

For stochastic effects, we note that the model can in principle be derived as the Fokker-Planck-system of a piecewise deterministic stochastic process (transitions at random points in time, deterministic evolution of immunity waning between transitions). In addition, diffusion terms could be introduced to model random variations in parameters or in the waning process. We leave these possible extension for future research.

Given its flexibility, the dual-age approach has distinct features that may be useful in public health decision making. It naturally accommodates
realistic parameter dependencies that reflect biological mechanisms: susceptibility genuinely depends on time since last exposure. The dynamics of the pathogen load depends on both the infection stage and the prior immunity, and vaccination policies can rationally target individuals based on the age of immunity rather than treating all susceptibles uniformly. The framework is flexible and allows for extensions such as demographic processes for long-term forecasting and immunity-age-dependent
vaccination strategies for policy evaluation. The explicit tracking of immunity age enables modeling of booster vaccination campaigns that target individuals whose immunity has waned beyond a specified threshold, a policy tool not feasible in standard epidemiological  compartmental models.
The ability to incorporate such nuanced vaccination policies demonstrates the flexibility of the dual-age
framework for capturing realistic intervention strategies.

\section*{Author contributions}
\begin{itemize}
    \item Raimund M. Kovacevic: Conceptualization, Formal analysis, Investigation, Methodology, Software, Writing - original draft and revision
    \item Nikolaos I. Stilianakis: Conceptualization, Methodology, Investigation, Validation, Writing -- original draft and revision
    \item Vladimir M. Veliov: Conceptualization, Formal analysis, Investigation, Methodology, Numerical method, software and simulations, Writing -- original draft and revision

\end{itemize}

\section*{Conflicting Interest}
Nothing to declare
\section*{Funding}
The last author is partly supported by 
the Bulgarian Ministry of Education and Science, Scientific Programme 
”Enhancing the Research Capacity in Mathematical Sciences (PIKOM)”,
No. DO1-67/05.05.2022.

The funding bodies had no role in the formulation of the model, the mathematical 
analysis, the computational methodology, or the interpretation of the numerical findings.
\section*{Disclaimer}
The views expressed are purely those of the writer (NIS) and may not in 
any circumstance be regarded as stating an official position of the European Commission.



\begin{thebibliography}{99}

\bibitem[Angelov et al. (2024)]{Angelov+2024}
Angelov G, Kovacevic RM, Stilianakis NI, Veliov VM. An immune-epidemiological model with waning immunity after infection or vaccination, Journal of Mathematical Biology, 88:71, 2024.

\bibitem[Barbarossa and Roest (2015)]{Barbarossa+Roest-2015}
Barbarossa MV, Röst G. Immuno-epidemiology of a population structured
by immune status: a mathematical study of waning
immunity and immune system boosting. {\em J Math Biol}  71:1737-1770, 2015. 

\bibitem[Barbarossa et al. (2017)]{Barbarossa+2017}
Barbarossa MV, Polner M, Röst G. Stability switches induced by immune system boosting in an SIRS model with discrete and distributed delays. {\em SIAM J Appl Math 77(3):905–923}, 2017. 

\bibitem[Barbarossa et al. (2018)]{Barbarossa+2018}
Barbarossa MV, Polner M, Röst G. Temporal evolution of immunity distributions in a population with waning and boosting. {\em Complexity} (1):9264743, 2018.

\bibitem[Bobrovitz et al. (2023)]{Bobrovitz+2023}
Bobrovitz N, Ware H et al.
Protective effectiveness of
previous SARS-CoV-2 infection and hybrid immunity against the omicron variant and severe disease: A systematic review and meta-regression. {\em Lancet Infect. Diseases}, 23, 556–567, 2023.

\bibitem[Burke et al. (2020)]{Burke2020}
Burke LG, Orav EJ, Zheng J, Jha AK.
Healthy Days at home: A novel population-based outcome measure.
\textit{Healthc (Amst)}, 8(1):100378, 2020.

\bibitem[Chamaitelly et al. (2025)]{Chamaitelly+2025}
Chemaitelly H, Ayoub HH, Coyle P, et al. Differential protection against SARS-CoV-2 reinfection pre- and post-Omicron. {\em Nature}, 639:1024-1031, 2025.

\bibitem[Diekmann et al. (2018)]{Diekmann+2018}
Diekmann O, Graaf W, Kretzschmar M, Teunis P. Waning and boosting: on the dynamics of immune status. {\em J. Math. Biology} 77:2023--2048, 2018.

\bibitem[Diekmann  and Heesterbeek (2000)]{Diekmann+H-00}
Diekmann O, Heesterbeek JAP. {\em Mathematical epidemiology of infectious diseases.} Wiley series of mathematical and computational biology. Wiley, England, 2000.

\bibitem[Diekmann  et al. (1990)]{Diekmann+H+M-90}
Diekmann O, Heesterbeek JAP, Metz J. On the Definition and the Computation of the Basic Reproduction Ratio R0 in Models For Infectious-Diseases in Heterogeneous Populations. {\em Journal of Mathematical Biology.} 28(4):365-82, 1990.

\bibitem[Fonville et al. (2014)]{Fonville+2014}
Fonville JM, Wilks SH, et al. Antibody landscapes after influenza virus infection or vaccination.
Science. 346(6212):996–1000, 2014

\bibitem[Heesterbeek and Dietz (1996)]{HeesterbeekDietz-96}
Heesterbeek JAP, Dietz K. The concept of $R_0$ in epidemic theory. {\em Statistica Neerlandia} 50(1):89-110, 1996.

\bibitem[Ghosh et al.(2023)]{Ghosh+Volpert-23}
Ghosh S, Volpert V, Banerjee M. 
An age-dependent immuno-epidemiological model with distributed recovery and death rates. {\em Journal of Mathematical Biology}, 86:21, 2023.

\bibitem[Ghosh et al.(2025)]{Ghosh+2025}
Ghosh S, Volpert V, Banerjee M. Estimation of time-varying recovery and death rates from epidemiological data: A new approach. {\em Mathematical Biosciences}, 387:109479, 2025.

\bibitem[Goldberg et al. (2022)]{Goldberg+2022}
Goldberg Y, Mandel M, et al.
Protection and waning of natural and hybrid immunity to SARS-CoV-2. {\em N Engl J Med} 386:2201–2212, 2022. 


\bibitem[Ke et al.(2022)]{Ke+-22}
Ke R et al. Daily longitudinal sampling of SARS-CoV-2 infection reveals substantial heterogeneity 
in infectiousness. {\em Nature Microbiology}, 7, 640--652, 2022. 
https://doi.org/10.1038/s41564-022-01105-z              

\bibitem[Khan and Tanimoto (2024)]{KHAN2024657}
Khan Md. MUR, Tanimoto J. Influence of Waning Immunity on Vaccination Decision-Making: {{A}} Multi-Strain Epidemic Model with an Evolutionary Approach Analyzing Cost and Efficacy. {\em Infectious Disease modeling}, 9(3):657--675, 2024.

\bibitem[Krammer (2019)]{Krammer-2019}
Krammer F. The human antibody response to influenza A virus infection and vaccination. {\em Nature Reviews Immunology} 19(6):383–397, 2019.

\bibitem[Lam et al. (2021)]{Lam-2021}
Lam MB, Riley KE, et al. 
Healthy days at home: A population-based quality measure for cancer patients at the end of life.
\textit{Cancer}, 2021.

\bibitem[Lindsey et al. (2025)]{Lindsey+2025}
Lindsey KM, Farrell Z et al.
From first infection to reinfection: Comparing nucleocapsid antibody kinetics in vaccinated and unvaccinated adults. {\em Vaccine}, 62:127593, 2025.

\bibitem[Nishimura et al. (2023)]{NISHIMURA2023113426}
Nishimura I, Arefin R, Tatsukawa Y, Utsumi S, Hossein A and Tanimoto J. 
Social Dilemma Analysis on Vaccination Game Accounting for the Effect of Immunity Waning.
\textit{Chaos, Solitons \& Fractals} 171:113426, 2023.

\bibitem[Okuwa et al. (2021)]{Okuwa+2021}
Okuwa K, Inaba H, Kuniya T. An age-structured epidemic model with boosting and waning of immune status. {\em Math Biosci Eng} 18(5):5707–5736, 2021.

\bibitem[Rao and Reimherr (2023)]{Rao+2023}
Rao, AR, Reimherr, M. Modern non-linear function-on-function regression. {\em Stat Comp} 33:130, 2023.

\bibitem[Sanches-de Prada et al.  (2024)]{Sanches_de_Prada+2024}
Sánchez-de Prada L, Martínez-García MA et al.
Impact on the time elapsed since SARS-CoV-2 infection, vaccination history, and number of doses, on protection against reinfection.
{\em Scientific Reports}, 14 (1), 353, 2024.

\bibitem[Schuh et al. (2024)]{LS+PM+NS+VV-24}
Schuh L, Markov PV, Veliov VM and Stilianakis NI.
A mathematical model for the within-host (re)infection dynamics of SARS-CoV-2. Mathematical Biosciences, 371, 109178, 2024.

\bibitem[Sharpe and Lotka (1911)]{Sharpe+1911}
Sharpe FR, Lotka AJ. A problem in age-distribution. {\em Philosophical Magazine} 21: 435-438, 1911.

\bibitem[Srivastava et al. (2024)]{Srivastava+2024}
Srivastava K, Carreno JM et al.
SARS-CoV-2-infection- and vaccine-induced antibody responses are long lasting with an initial waning phase followed by a stabilization phase. {\em Immunity}, 57:587-599, 2024.

\bibitem[Statistik Austria (2025)]{StatistikAustria-AMDC}
Statistik Austria. Austrian Micro Data Center (AMDC) -- COVID-19 Impf- und Infektionsdaten. 
Available at: https://www.statistik.at/amdc-data/\#/product. Accessed 9 February 2025.


\bibitem[Vattiato et al. (2022)]{Vattiato+2022}
Vattiato G, Lustig A, Maclaren OJ, Plank MJ, modeling the dynamics of infection, waning immunity and re-infection with the Omicron variant of SARS-CoV-2 in Aotearoa New Zealand. {\em Epidemics}, 41:100657, 2022.

\bibitem[Veliov (2008)]{VV:JMAA_08}
Veliov VM.
Optimal Control of Heterogeneous Systems: Basic Theory.
{\em J. Math. Anal. Appl.}, 346:227--242, 2008.		


\bibitem[Wolthuis et al. (2017)]{Wolthuis+2017}
Wolthuis RG, Wallinga J, van Boven M. Variation in loss of immunity shapes influenza epidemics and the impact of vaccination, {\em BMC Inf. Dis.}, 17:632, 2017.

\bibitem[Yang et al. (2024)]{Yang+2024}
Yang  J, Duan X, Sun G. An immune-epidemiological model with non-exponentially distributed disease stage on complex networks. {\em J Theor. Biol.} 595:111964, 2024.

\bibitem[Yang and Nakata (2021)]{Yang+Nakata-2021}
Yang L, Nakata Y. Note on the uniqueness of an endemic equilibrium of an epidemic model with boosting of immunity. {\em J Biol Syst} 29(02):291–302, 2021.

\end{thebibliography}
\end{document}